\documentclass[aps,prb,amsmath,amssymb,floats,twocolumn,showpacs]{revtex4}

\usepackage[dvips]{epsfig}

\begin{document}
 \title{Nonequilibrium transfer and decoherence in quantum
impurity problems}
 \author{Holger Baur,$^1$ Andrea Fubini,$^2$ and Ulrich Weiss$^1$}
 \affiliation{$^1$Institut f\"ur Theoretische Physik, Universit\"at Stuttgart,
   D-70550 Stuttgart, Germany}
 \affiliation{$^2$Dipartimento di Fisica, Universit\`a di Firenze,
    and INFM,  Via G.Sansone~1, I-50019 Firenze, Italy}
 \date{\today}

 \begin{abstract} 
Using detailed balance and scaling properties of integrals that appear in the Coulomb gas 
reformulation of quantum impurity problems,  we establish exact relations between the
nonequilibrium transfer rates of the boundary sine-Gordon and the anisotropic Kondo model 
at zero temperature.  Combining these results with findings from the thermodynamic Bethe 
ansatz, we derive exact closed form expressions for the transfer rate in the biased 
spin-boson model in the scaling limit.  They illustrate how the crossover from weak to 
strong tunneling takes place.  Using a conjectured correspondence between the transfer and
the decoherence rate, we also determine the exact lower bound for damping
of the coherent oscillation as a function of bias and dissipation strength in this 
paradigmic model for NMR and superposition of macroscopically distinct states (qubits).  
\end{abstract}
 \maketitle

\section{Introduction}\label{secintro}
Quantum impurity problems (QIPs) have attracted a great deal of interest 
recently. This is because the underlying physics is non-trivial and the models
are manageable technically despite their essentially nonperturbative nature.
In addition, they have a multitude of experimental applications, including
the Kondo effect, quantum dots, dissipative quantum mechanics, tunneling in
quantum wires and fractional quantum Hall devices.\cite{weissbook} 

There have been discovered various relations between 
thermodynamic quantities of the 
anisotropic Kondo model (AKM) and the boundary sine-Gordon (BSG) field theory
model.\cite{fendley95a} 
Each of these integrable models is of considerable interest and it is 
remarkable that they are closely related.
They are both boundary integrable quantum field theories with a quantum-group spin at
the boundary which takes values in standard or cyclic representations of the quantum group
$SU(2)_q$.\cite{fendley95c}
 
The AKM is related to the spin-boson (SB) model.~\cite{leggett87} The BSG model is equivalent to 
the Schmid model, which describes a damped particle in a tilted cosine potential~\cite{schmid83}
and has a great variety of applications to charge and particle transport
including Josephson junction dynamics.~\cite{weissbook}
The field theory limit in the AKM and BSG model corresponds to the Ohmic scaling limit in 
the SB  and Schmid model. 
Here we are interested in the strong-backscattering
or tight-binding (TB) representations of these models.
The equivalence and difference of the various models can most easily be seen
in the Anderson-Yuval Coulomb gas representation for the partition
function, which is the grandcanonical sum of a one-dimensional
classical gas of positive and negative unit charges with ''log-sine''
interactions and overall neutrality. The AKM or SB model differs
from the BSG or Schmid model by a different charge ordering.
Principally in the sequel, we concentrate on the Schmid and SB model and use the language of 
dissipative quantum mechanics. With use of the correspondence relations, our results directly 
apply also to the BSG and the anistropic Kondo model. 

Here we study nonequilibrium quantum dynamics of the models mentioned
with the rigorous Keldysh approach. Upon uncovering functional relations between 
perturbative Coulomb integrals, we discover that at zero temperature for every particular 
transfer rate only paths with minimal number of
tunneling moves contribute. This allows us then to find the characteristic function encapsulating
all statistical fluctuations of the 
transport process. The results for the Schmid model 
corroborate the findings from the thermodynamic Bethe ansatz.
We discover, as a second result of current interest, an intriguing functional relation
between transfer rates in the Schmid and partial transfer rates in the SB model. The major result 
is given in Eq.~(\ref{raterel}). Using this, we derive a concise
contour integral expression for the transfer rate in the SB model [cf. Eq.~(\ref{intrep1})], 
from where we find both the weak- and the strong-tunneling expansion for this rate.
We also make and check a conjecture which relates the transfer rate to the
decoherence rate in the SB model. The ensuing asymptotic series, Eq.~(\ref{deco}), provides an 
exact lower bound for decoherence in the scaling limit as a function of bias and damping 
strength.

In the following section, we briefly review a general Poissonian transport model 
with transfer rates  describing direct forward and backward transitions by $m$ states,
$m=1,\,2,\,3,\cdots$. We solve the master equation and determine the evolution of the 
probability distribution. In Sec.~\ref{secnonequ}, we study the Schmid model in the
real-time Coulomb gas representation, which is based on the nonequilibrium Kel\-dysh 
method. Upon performing a cluster decomposition,
we show that the BSG dynamics at long times represents a realization of the Poissonian
dynamics studied in Sec.~\ref{secpoisson}. We derive exact formal expressions for the 
quantum transfer rates in which friction and noise effects are clearly separated.
The corresponding investigation for the AKM or SB model is presented in Sec.~\ref{secinc}. 
The weak-tunneling expansion for the transfer rates at zero temperature are studied
both for the Schmid and SB model in Sec.~\ref{secwt}. Using detailed
balance relations, and scaling properties holding for a subgroup of the Coulomb integrals,
we uncover a multitude of linear functional relations between the occuring
Coulomb integrals of same order. Using these relations, we discover formidable cancellations
in the path sum for the individual rates of the Schmid model such that
only the paths with minimal number of
tunneling moves contribute. This peculiarity allows us to determine
all statistical fluctuations once the current is known. The corresponding discussion is
given in Subsec.~\ref{secfluc}. We also discover an intriguing functional relation between
the rates in the Schmid model and the partial rates in the SB model.
The findings in Sec.~\ref{secwt} can be used to shed light on the entire regime from weak to 
strong tunneling, both in the Schmid and SB model.
In Sec.~\ref{secwtos}, we derive from the perturbative series an
exact integral representation for the transfer rate in the
SB model at zero temperature valid for general coupling, bias and damping strengths. 
From that, the asymptotic
(strong-tunneling) expansion for the transfer rate is derived. 
In Subsec.~\ref{secconj}, we propose a conjecture about a simple close relation between the
relaxation and decoherence rate in the SB model. The conjectured formula for the
decoherence rate is in agreement with all known special cases and gives a lower bound 
on decoherence for general bias and damping strength. Finally, in Subsec.~\ref{subsecenhanc}
we give analytic expressions for the leading thermal corrections.

\section{Poissonian transport and noise}\label{secpoisson}

Consider a general model describing transport of mass or charge
between sites $ n=0,\, \pm 1,\,\pm 2, \cdots$ via
 direct forward and backward transitions by $\ell$ sites,
$ \ell = 1,\, 2,\cdots $. The respective weights per unit time (transfer ``rates'') are 
denoted by $\gamma_{\ell}^{\pm}$. Assuming statistically independent transitions, the
dynamics of the population probability $P_n(t)$ of site $n$ is
governed by the master equation
\[
\dot{P}_n(t) = \!\sum_{\ell=1}^\infty \left[ \gamma_\ell^+ P_{n-\ell}(t) + \gamma_\ell^- 
P_{n+\ell}(t) - (\gamma_\ell^+\! +\! \gamma_\ell^-) P_n(t)\right]\, .  
\]
The Fourier transform 
$\tilde{P} (k,t) = \sum_n e^{ikn}_{}\, P_n(t)$  with initial state
$P_n^{}(0) = \delta_{n,0}^{}$ is found from this equation as 
\[
\tilde{P} (k,t) = 
\prod_{n=1}^\infty \exp\left[t\left(e^{ikn}_{} -1\right)\gamma_n^+
+ t\left(e^{-ikn}_{} -1\right)\gamma_n^- \right] .
\]
The moments of the probability distribution follow from the characteristic function
$\tilde{P} (k,t)$ by differentiation,
\begin{equation}
\label{moments1}
	\left< n^m (t) \right> \equiv \sum_n n^m P_n(t) =
 \left. 
  \left( -i \frac{\partial}{\partial k} \right)^m 
\tilde{P}(k,t) \right|_{k=0} \;. 
\end{equation}
The resulting expressions can be written in terms of the irreducible moments or 
cumulants $\left< n^m(t) \right>_c$ as
\begin{eqnarray}\nonumber
  \left< n(t) \right> &=&  \left< n(t) \right>_c \;,\\  \label{moments2}
  \left< n^2(t) \right> &=&
                  \left< n(t) \right>_c^2 + \left< n^2(t) \right>_c \;, \\ 
	\left< n^3(t) \right> &=& \left< n(t) \right>_c^3 + 3 \left< n^2(t) \right>_c 
		\left< n(t) \right>_c + \left< n^3(t) \right>_c \;.  \nonumber
\end{eqnarray}
The cumulants expansion
\begin{equation}\label{cumexpan}
  \ln \tilde{P}(k,t) = \sum_{m=1}^\infty \frac{(ik)^m}{m!} \left< n^m(t)\right>_c 
\end{equation}
leads us to the expressions
\begin{equation}
\begin{array}{rcl}
\left< n^m(t)\right>_c &=& {\displaystyle
\left. \left( -i \frac{\partial}{\partial k} \right)^m 
\ln \tilde{P}(k,t) \right|_{k=0}} \\[3mm] 
       &=& {\displaystyle
	t \sum_{n=1}^\infty n^m_{} \left[ \gamma_n^+  + (-1)^m \gamma_n^-  \right] \;.   }
\end{array}
\end{equation}

The function $\tilde{P}(k,t)$ may be conveniently rewritten in terms 
of partial forward/backward currents $I_n^\pm = n \gamma_n^\pm $ as
\begin{eqnarray}\nonumber
\tilde{P}(k,t) &=& {\displaystyle \prod_{n=1}^\infty Z_n^+(k,t)\, Z_n^-(k,t) , } \\
Z_n^\pm (k,t) &=& {\displaystyle
\sum_{\ell=0}^\infty e^{ikn\ell}_{} \frac{(tI_n^\pm/n)^\ell}{\ell!}e^{-tI_n^\pm/n}  }
\label{charfunc1} \\    \nonumber 
&=& {\displaystyle \exp [t(e^{ikn}-1)I_n^\pm/n\,] \; . }     
\end{eqnarray}

The characteristic function $\tilde{P}(k,t)$ encapsulates all statistical
fluctuations of the transport process.  
To elucidate the physical meaning of the above expressions, suppose now that
the mass transferred per unit time in forward direction were the result of a Poisson 
process for particles of unit mass propagating via uncorrelated nearest-neighbour
forward steps, contributing a current $I_1^+ = \gamma_1^+$, plus a Poisson process
of uncorrelated forward moves via next-to-nearest-neighbour transitions contributing
a current $I_2^+ = 2 \gamma_2^+$, etc. Further suppose that the total backward 
current were the result of independent Poisson processes with corresponding 
partial backward currents $I_n^-$, $n=1,2,\cdots$. The final form of the 
characteristic function would then be the expression (\ref{charfunc1}).

With regard to applications of this model to QIPs (see below), there is a different 
interpretation. Expression (\ref{charfunc1}) may also
represent independent Poisson processes for particles of charge one going
across the impurity's barrier in forward/backward direction, contributing
a current $I_1^\pm$, plus a Poisson process for particles of charge two (or for a pair
of particles of charge one) contributing a forward/backward current $I_2^\pm$, etc.

Next, suppose that the particle is coupled to a thermal reservoir at temperature $T$. 
Then the forward and backward transfer weights are related by detailed balance.
Assuming that the potential drop per site interval in forward direction is
$\epsilon> 0$ (we use units where $\hbar = k_{\rm B}^{} =1$), we then have
\begin{equation}\label{detbal}
 \gamma_n^{-} = e^{-n\epsilon/T} \gamma_n^{+}  \; . 
\end{equation}

For classical Poisson processes, all of the transfer weights $\gamma_n^\pm$,
$n=1,\,\cdots$, must be positive. When the Poisson processes come about 
quantum-mechanically, the transfer weights may be partly negative (see below).
This does not spoil conservation of probability, since $\sum_n
\dot{P}(t) =0 $ by construction of the master equation regardless of
the forms chosen for the set $\{\gamma_n^{\pm}\}$.

\section{\label{sec3} Nonequilibrium quantum transport in the 
BSG or Schmid model}\label{secnonequ}

The nonequilibrium transport through a local backscattering potential embedded in a 
Luttinger liquid environment gives a fingerprint of the non-Fermi-liquid state.
This situation is realized e.g. in resonant tunneling experiments through
a point contact in quantum Hall devices.\cite{roddaro,picciotto} 
In the field theory limit, the tunneling 
problem is described by the BSG model, which is a harmonic bosonic field on a half-line, 
and the only interactions take place on the boundary.~\cite{fendley95c} 
The Hamiltonian for strong-backscattering is $H_{\rm BSG}^{} = H_0^{} + H_{\rm Bound}^{}$, where
\begin{eqnarray}\nonumber
H_{0}^{} &=& {\displaystyle
  \frac{g}{4\pi}\int_0^\infty \!\!\!\!{\rm d}x  \left[\, \Pi_{}^2 
 + (\partial_x^{} \phi)^2_{} \,\right] +  \frac{V}{2\pi}\! 
\int_0^\infty \!\!\!{\rm d}x\, \partial_x^{}\phi \;, } \\[1mm]  \label{hambsg} 
 H_{\rm Bound}^{} &=& -\; {\displaystyle 
\;  \Delta_{\rm S}^{}\, \cos[\,\phi(0)\,] } \;.
\end{eqnarray}
Here, $H_0^{}$ represents the bulk and an applied voltage $V$, and $H_{\rm Bound}^{}$ is the
interaction on the boundary.

The Schmid model describes a quantum Brownian particle moving in a tilted cosine 
potential.~\cite{schmid83} In the TB representation, the Hamiltonian is (except for a counterterm)
\begin{equation}\label{schmidmod}
\begin{array}{rcl}
H_{\rm S}^{} &=& - {\displaystyle \frac{\Delta_{\rm S}^{}}{2}\! \sum _n \left( a_n^\dagger 
a_{n+1}^{} + {\rm h.c.} \right) 
\, - \, \epsilon \sum_n n\,a_n^\dagger a_n^{}  }    \\[4mm] 
 &&  -\; {\displaystyle \sum_\alpha c_\alpha^{} x_\alpha^{} 
\sum_n na_n^\dagger a_n^{} \, + \, H_{\rm bath}^{}(\,\{x_\alpha^{}\}\,) }  \, .
\end{array}
\end{equation}
The first line gives the TB system. Here, $\Delta_{\rm S}^{}$ is the coupling 
energy and $\epsilon$  is the potential drop or bias energy 
between neighbouring sites (lattice constant $a =1$).
The second line describes the system-bath coupling and the harmonic bath.
In the Ohmic scaling limit, the spectral density 
$J(\omega)= \pi\sum_\alpha \delta(\omega - \omega_\alpha^{}) 
c_\alpha^2/(2 m_\alpha^{} \omega_\alpha^{})$ is strictly linear in $\omega$,
$J(\omega) = 2K\omega$. The Kondo parameter $K$ is a 
dimensionless Ohmic damping strength,~\cite{kondo88}
and is the inverse of the Luttinger parameter $g$
or the filling fraction $\nu$ in fractional quantum Hall systems.

The effects of the Luttinger liquid bulk are in the two-point function on the boundary,
$\langle e^{i\phi(0,t)}_{}\,e^{-i\phi(0,0)}_{}\rangle_T^{} = e^{-Q(t)}_{}$. The function 
$Q(t)= \langle\, [\phi(0,0)-\phi(0,t)]\,\phi(0,0)\rangle_T^{}$ directly corresponds to the thermal
correlator in the Schmid model.
Introducing a high-frequency cutoff $\omega_c^{}$, we get in the field theory or scaling limit
\begin{equation}\label{inter2}
Q(t) = 2K \ln[\,(\omega_{\rm c}/\pi T)\sinh(\pi T |t|)\,] 
+ i\,\pi K\,{\rm sgn}(t) \;.
\end{equation}
The BSG model is in correspondence with the Ohmic Schmid model if we put $K=1/g$ and 
$\epsilon = 2\pi V$.
The system (\ref{hambsg}) or (\ref{schmidmod}) represents a quantum-mechanical 
realization of the general model discussed in Section II.

The nonequilibrium transport may be found upon employing the Feynman-Vernon or 
Keldysh formalism to the calculation of the reduced density matrix (RDM).
It is convenient to parametrize the sudden moves of the system along the forward 
path $q$ and backward path $q'$ in terms of charges $\{ u_j =\pm 1\}$ and
$\{ v_i = \pm 1 \}$, respectively.
It is then straightforward to derive the exact series expansion in the number of
tunneling transitions for the characteristic function $\tilde{P}(k,t)$. This series
has formal similarity to the partition function of charges in a one-dimensional
fixed volume. Taking the Laplace transform $\hat{P}(k,\lambda)$ formally corresponds
to switching to the analog of an isobaric ensemble of charges.
Upon tracing out the thermal bath, the charges are interacting
with each other through the complex time-nonlocal interaction $Q(t)$. 

Consider an individual double path or charge sequence with $n + 2\ell$ time-ordered 
$u$ charges on the $q$ path at locations $t_1 \le t_2\cdots\le t_{n+2\ell}$
and $n + 2 m $ time-ordered $v$ charges on the $q'$ path
at locations  $t'_1\le t'_2 \cdots \le t'_{n+2m}$, 
\begin{eqnarray}\label{clus}
\alpha_{\ell,m}^{\pm n} &\equiv& \{ q_\ell^{\pm n} ;\; {q'}_m^{\pm n} \}  \\
&\equiv& \{u_1,u_2\cdots,u_{n+2\ell};\; v_1,v_2,\cdots,v_{n+2m}\}_{\pm n} \; ,
\nonumber
\end{eqnarray}
where the total charges satisfy the constraints
\begin{equation}\label{constraint1}
\sum_{j=1}^{n+ 2\ell} u_j = \sum_{i=1}^{n+ 2m} v_i = \pm n   \; .
\end{equation}
The tunneling amplitude factor for this arrangement is
\begin{equation}\label{amplfac}
A[\alpha_{\ell,m}^{\pm n}] = \left(\frac{i\Delta_{\rm S}^{}}{2}\right)^{n+2\ell}
                            \left(\frac{-i\Delta_{\rm S}^{}}{2}\right)^{n+2m} \; ,
\end{equation}
and the bias phase factor reads 
$B[\alpha_{\ell,m}^{\pm n}] = e^{i\varphi[\alpha_{\ell,m}^{\pm n}]}_{}$ with 
\begin{equation}\label{biasphase}
\varphi[\alpha_{\ell,m}^{\pm n}] = {\textstyle
\epsilon\left( \sum_i v_i^{} t'_i - \sum_j u_j^{} t_j^{}  \right)  }   \; .
\end{equation}
The interactions for the arrangement (\ref{clus})  are described
by the Gaussian influence function 
\begin{equation}
{\cal F}[ \alpha_{\ell,m}^{\pm n}] = {\cal F}_1[q_\ell^{\pm n}]\,
{\cal F}_1^\ast[{q'}_m^{\pm n}]
\,{\cal F}_2[q_\ell^{\pm n},{q'}_m^{\pm n}]\;,
\end{equation}
where ${\cal F}_1$ and ${\cal F}_1^\ast$ describe the selfinteractions of the $q$ and
$q'$ path, respectively, and ${\cal F}_2$ the mutual interactions,
\begin{eqnarray}\nonumber
 {\cal F}_1[q_\ell^{\pm n}] &=& 
\exp\bigg\{\sum_{j>i =1}^{n+2\ell} u_j u_i Q(t_j - t_i)\bigg\} \; , \\ \label{infl}
{\cal F}_1^\ast[{q'}_m^{\pm n}] &=&  
\exp\bigg\{  \sum_{j>i=1}^{n+2m} v_j v_i Q^\ast (t_j' - t_i') 
\bigg\} \; , \\
{\cal F}_2[\alpha_{\ell,m}^{\pm n}] &=& 
\exp\bigg\{ - \sum_{j=1}^{n+2m} \sum_{i=1}^{n+2\ell} v_j u_i Q(t_j' -t_i) 
   \bigg\}\; .   \nonumber
\end{eqnarray}

The Poissonian dynamics (\ref{charfunc1}) appears at times $t$ large compared to the
transition times between the sites.
It is found upon performing a cluster decomposition of the individual charge arrangements 
contributing to $P_n^{}(t)$. The clusters represent the elementary processes of the tunneling 
dynamics. They are charge sequences corresponding to irreducible path sections 
which start and end in diagonal states of the RDM. By definition, clusters are noninteracting.
Therefore in Laplace space, a succession
of clusters factorizes and every interval in a diagonal state separating neighbouring 
clusters generates an additional factor of 
$1/\lambda$. Path segments which have interim visits of diagonal states can be divided up into
a reducible part, which breaks down into a succession of clusters of lower order, and an
irreducible remnant (see below). 

When the Laplace variable $\lambda$ goes to zero, the clusters become independent of $\lambda$.
As a result, the dynamics turns into Markovian behaviour at long times. 
In this limit, the totality of irreducible clusters describing the elementary processes 
$0 \to \pm n$ and $0 \to 0$ represent the weight per unit time $\gamma_n^{\pm}$ for direct
transitions from arbitrary site $m$ to site $m \pm n$, and the weight per unit time
$\gamma_0^{}$ to stay at the same site, respectively. 
Since the system with initial state
$P_n^{}(0)= \delta_{n,0}^{}$  must finally take real occupation of site $n$, there is an 
excess of forward moves over the backward moves for $n>0$. Thus we find from the 
Coulomb gas or path sum (we put $\delta(m,n) =\delta_{m,n}^{}$)
\begin{eqnarray}\nonumber
P_n^{}(t) &=& e^{\gamma_0^{} t}_{}\,\prod_{i=1}^\infty \Bigg\{
\sum_{j_i^{}=0}^\infty\; \sum_{\ell_i^{}=0}^\infty\;
\frac{(\gamma_i^+ t )^{j_i^{}}_{}\,(\gamma_i^- t)^{\ell_i^{}}}{j_i^{}!\,\ell_i^{}!}\Bigg\}
\\ \label{popul}
&& \quad\times\;\delta\left(\sum_{i=1}^\infty i (j_i^{} -\ell_i^{}),\, n \right) \;. 
\end{eqnarray}

The transfer ``rate'' $\gamma_n^{+}$ can be divided up into 
$\gamma_n^{+}(+)$ and $\gamma_n^{+}(-)$, in which the last charge is positive
and negative, respectively, $\gamma_n^+ = \gamma_n^{+}(+) + \gamma_n^{+}(-)$. 
Next, assume that we
replace the last charge in $\gamma_n^{+}(-)$, which is either $u_{\rm f}^{}=-1$ or
$v_{\rm f}^{}=-1$, by the charge $v_{\rm f}^{}=+1$ or $u_{\rm f}^{}+1$.
Under this substitution,  the bias phase and the influence factor are left unchanged
while the amplitude (\ref{amplfac})
changes sign, and the constraint  $n$ in Eq.~(\ref{constraint1}) changes into $n +1$,
as we see from Eqs.~(\ref{constraint1}) - (\ref{infl}). 
Therefore we have $\gamma_n^{+}(-) = - \gamma_{n+1}^{+}(+)$. 
Similarly, we get $\gamma_n^{-}(+) = - \gamma_{n+1}^{-}(-)$.
Thus we find $\gamma_n^\pm = \gamma_n^\pm(\pm) - \gamma_{n+1}^\pm(\pm)$ and 
$\gamma_0^{} = -\gamma_1^{+}(+) - \gamma_1^{-}(-)$, and therefore finally
\begin{equation}\label{normcond}
\gamma_0^{} + \sum_{n=1}^\infty (\gamma_n^{+} +\gamma_n^{-}) = 0 \; .
\end{equation}  
This relation ensures conservation of probability of the expression (\ref{popul}),
$\sum_n P_n^{}(t) =1$. Furthermore, the characteristic function $\tilde{P}(k,t)$ is found 
to take the form (\ref{charfunc1}). 

The irreducible contribution of the charge arrangement (\ref{clus})
with excess charge $\pm n$ in Eq.~(\ref{constraint1}) gives a partial contribution to
the transfer ``rate'' $\gamma_n^\pm$. We find
\begin{eqnarray}\label{partial1}
\gamma_n^\pm[\alpha_{\ell,m}^{\pm n}] &=&  {\displaystyle
(-1)^{\ell + m}\left(\frac{\Delta_{\rm S}^{}}{2}\right)^{2(n+\ell+m)} 
\int_{-\infty}^{\infty} \!\! {\rm d}\tau } 
\\[4mm]  \nonumber
&\times& {\displaystyle  \!\!\!
\int\limits_0^\infty \;\prod_{j=1}^{n+2\ell -1}\!\!\!{\rm d}\rho_j \!\!\! \prod_{i=1}^{n+2m -1} 
\!\!\! {\rm d}\rho'_i  \,\,e^{i \varphi[\alpha_{\ell,m}^{\pm n}]}_{}\,
{\cal F}^{(c)}_{}[ \alpha_{\ell,m}^{\pm n}]} \;,  
\end{eqnarray}
where ${\cal F}^{(c)}$ is the irreducible
part of the influence function. The integrations are over the charge 
intervals $\rho_j = t_{j+1} -t_j$ and $\rho'_i = t'_{i+1} -t'_i$. The additional
unrestricted integration over the interval $\tau = t_1' -t_{n + 2\ell}$ introduces 
$N_{n;\ell,m}=\frac{(2n + 2\ell + 2m)!}{(n+2\ell)!\,(n+2m)!}$ different
possibilities of mixing up the time-ordered moves in $q_\ell^{\pm n}$ with those in
${q'}_m^{\pm n}$. In general, the partial transfer weight (\ref{partial1}) is
complex. The complex conjugate counterpart comes
from the set (\ref{clus}) in which the $u$ and $v$ charges are 
interchanged. 

For given $n$, the term with $\ell=m=0$ in Eq.~(\ref{partial1}) has
minimal number of tunneling moves. The contributions with
$\ell, m >0$ have additional pairs of forward/backward moves both along the
path $q$ and the path $q'$.
The grand-canonical ensemble of charge arrangements satisfying the constraint
(\ref{constraint1}) yields the transfer rate $\gamma_n^\pm$,
\begin{equation}
\gamma_n^\pm = \sum_{\ell =0}^\infty \sum_{m=0}^\infty \sum_{\alpha_{\ell,m}^{\pm n}}
\gamma_n^\pm[\alpha_{\ell,m}^{\pm n}] \; .
\end{equation}
The sum $\sum_{\alpha_{\ell,m}^{\pm n}}$ is over the
$M_{n;\ell,m}= \frac{(n+2\ell)!}{(n+\ell)!\,\ell!}\, \frac{(n+2m)!}{(n+m)!\,m!}$ 
different possibilities to order the $u$ charges for fixed $n$ and $\ell$ and 
the $v$ charges for fixed $n$ and $m$.

The partial transfer weight (\ref{partial1}) takes up the tunneling 
contributions to the incoherent rate $\gamma_n^\pm$, in which (i) the succession of 
moves in the $q$ and $q'$ path is kept fixed separately, and (ii) all 
possible arrangements relative to each other are contained, the latter being due to the unrestricted
$\tau$-integration in Eq.~(\ref{partial1}).   
It describes a coherent tunneling process from site 0 to site $\pm n$
and takes into account $M_{n;\ell,m}$ different possibilities of visiting virtually
$2(n+\ell+m)-1$ intermediate states of the RDM.

It is straightforward to see that the partial ``rate'' (\ref{partial1}) obeys
detailed balance already. To this, observe that the expression (\ref{inter2}) is analytic 
for complex time $z$ in the strip $0 \ge {\rm Im}\,z > - 1/T $ and has the property
$Q(t-i/T ) = Q^\ast(t) = Q(-t)$. Thus we may shift the flip times of the backward 
path as $t'_i \to t_i' -i\bar{\tau}$, where $0\le \bar{\tau} <1/T $. This results
in a shift of the $\tau$ integration path in (\ref{partial1}),
$\tau \to \tau-i\bar{\tau}$. 
Since this path is unrestricted, the edge contours perpendicular to
the real-time axis do not contribute. 
Hence the integral does not depend on the shift $\bar{\tau}$. 
Putting $\bar{\tau}=1/T$, the bias factor (\ref{biasphase}) in Eq.~(\ref{partial1}) 
then generates a factor $e^{\pm n\epsilon/T}$ and the interaction term ${\cal F}_2$
transforms into ${\cal F}_2^\ast$, while ${\cal F}_1^{}$ and ${\cal F}_1^{\ast}$ are left 
unchanged. Next observe that under combination of time reversal and charge conjugation 
both the bias factor $e^{i\varphi}$ and the interaction 
factors ${\cal F}_1$ and ${\cal F}_1^\ast$ are invariant, 
while ${\cal F}_2^*$ transforms back into ${\cal F}_2$.
Thus we find
\begin{equation}\label{partdetbal}
\gamma_n^- [\alpha_{\ell,m}^{-n}] = e^{-n\epsilon/T} \, 
\gamma_n^+[\widetilde{\alpha}_{\ell,m}^{+n}] \;,
\end{equation}
where $\widetilde{\alpha}_{\ell,m}^{\mp n}$ is the charge sequence resulting 
from (\ref{clus}) by charge conjugation and time reversal.

In the above we have shown that the $\{u,v\}$ representation is useful to identify 
partial sums of tunneling paths for which detailed balance holds already. 
On the other hand, the disadvantage of this representation is that it
mixes up friction and noise. 
These effects can be separated clearly in the Ohmic scaling limit using a different
representation. To this end,
we now introduce charges $\eta_j =\pm 1$ describing
forward/backward moves along the quasiclassical path $q+q'$ and charges $\xi_i =\pm1$
representing sudden moves along the quantum fluctuation path $q-q'$. 
We then have the correspondence
\begin{eqnarray*}
u = \pm 1\qquad & \longleftrightarrow & \qquad \{\eta,\,\xi\} = \{\pm 1,\, \pm 1 \}\;, \\
v  = \pm 1\qquad & \longleftrightarrow & \qquad \{\eta,\,\xi\} = \{\pm 1,\, \mp 1 \}\; .
\end{eqnarray*}
The cumulative charges 
\begin{equation}\label{cumcha}
r_j^{} = \sum_{k=1}^j \eta_k^{}\quad\qquad\mbox{and} \quad\qquad p_j^{} = \sum_{k=1}^j \xi_k^{}
\end{equation}
measure propagation of the system after $j$ moves
in forward/backward and off-diagonal direction, respectively.

An individual tunneling path with $2\ell$ moves contributing to the rate 
$\gamma_n^\pm$, where $\ell\ge n$ is then parametrized by a set
of $\{ \eta,\xi \}$ pairs at locations $ t_1 \le t_2\le\cdots\le t_{2\ell}$, 
\begin{equation}\label{clus2}
\beta_{2\ell}^{\pm n} \equiv \{\eta_1,\xi_1;\,\eta_2,\xi_2;\,\cdots;\,\eta_{2\ell},
\xi_{2\ell}\,\}_{\pm n}\; ,
\end{equation}
where the charge arrangements are constrained as
\begin{equation}\label{constraint2}
\sum_{i=1}^{2\ell} \eta_i^{}  = \pm 2n \; ; \qquad \sum_{i=1}^{2\ell} \xi_i^{} = 0  \; .
\end{equation}

Next, we introduce dimensionless temperature $\vartheta=T/\epsilon$ and
dimensionless charge intervals 
$\tau_i^{} =\epsilon\,(t_{i+1} -t_i)$ where $i= 1,\cdots,2\ell-1$.
Then the irreducible influence factor for the charge sequence (\ref{clus2}) 
may be written as~\cite{weissbook}
\begin{equation}\label{inflfac}
{\cal F}^{(c)}_{}[\beta_{2\ell}^{\pm n}] = (\epsilon/\omega_c)^{2K\ell}_{}\,
G_\ell^{(c)}(\vec{\tau},\,\vec{\xi}\,)\, 
e^{- i\phi_\ell^{}(\vec{\eta},\,\vec{\xi}\,)}_{} \; .
\end{equation}
The influence phase $\phi_\ell^{}(\vec\eta,\vec{\xi}\,)$ adds up the imaginary parts 
of the bath correlations and describes friction,
\begin{equation}
\phi_\ell^{}(\vec{\eta},\vec{\xi}\,) = \pi K \sum_{j=1}^{2\ell-1} \eta_j^{}p_j^{}\; .
\end{equation}
Observe that $\phi_\ell$ depends only on the order of the charges, but not on 
the individual charge intervals $\tau_1^{},\,\cdots,\,\tau_{2\ell-1}^{}$.

The amplitude $G_\ell^{(c)}(\vec{\tau},\vec{\xi}\,)$ includes the quantum fluctuations.
The term $G_\ell^{\rm (subtr)}(\vec{\tau},\,\vec{\xi}\,)$ subtracted from 
$G_{\ell}^{}(\vec{\tau},\,\vec{\xi}\,)$,  
\begin{equation}\label{gsubtrac}
G_\ell^{(c)}(\vec{\tau},\,\vec{\xi}\,) = G_\ell^{}(\vec{\tau},\,\vec{\xi}\,) -
G_\ell^{(\rm subtr)}(\vec{\tau},\,\vec{\xi}\,) \;, 
\end{equation}
is to eliminate the reducible parts of $G_\ell^{}(\vec{\tau},\vec{\xi}\,)$,
if any. Introducing the distance of charge pair $\{j,i \}$,
\[
\tau_{ji}^{} = \epsilon\,( t_j - t_i) = \sum_{k=i}^{j-1} \tau_k^{}\; ,
\] 
the amplitude takes the form
\begin{equation}\label{noiseampl}
G_{\ell}(\vec{\tau},\vec{\xi}\,) = \exp\left(2K \!\!\sum_{j>i=1}^{2\ell}
\ln\left[\frac{\sinh(\pi \vartheta \tau_{ji}^{}) }{\pi 
\vartheta }\right]^{\xi_j \xi_i}  \right)         \; .
\end{equation}
Subtractions are necessary when a cluster splits up into neutral
subclusters, i.e. when one or several of the 
cumulative $p$-charges are zero. By definition, in the subtracted term 
$G_\ell^{\rm (subtr)}(\vec{\tau},\vec{\xi}\,)$ all interactions across an interval associated 
with zero cumulative charge are disregarded.
Hence, $G_\ell^{\rm (subtr)}(\vec{\tau},\vec{\xi}\,)$ 
factorizes into a product of amplitudes of lower order representing the subclusters.
Putting, e.g.,  $\ell =2$ and 
$\xi_2 = - \xi_1$, i.e. $p_2^{} =0$, we have
\[ 
G_{2}^{\rm (subtr)}(\vec{\tau},\vec{\xi}\,) 
=  \left(\frac{\pi \vartheta}{\sinh(\pi\vartheta \tau_1^{})}
\right)^{2K}\,\left( \frac{\pi\vartheta}{\sinh(\pi \vartheta \tau_3^{})}
\right)^{2K}_{}\; .
\]
The irreducible amplitude $G_\ell^{\rm (c)}(\vec{\tau},\vec{\xi}\,)$ 
encapsulates the interactions of the $\xi$ charges and
acts as a Gaussian filter controlling the quantum fluctuations. 

The residual factor $(\epsilon/\omega_c)^{2K\ell}$ in Eq.~(\ref{inflfac}) is an 
adiabatic 
Franck-Condon factor leading to renormalization of the tunneling amplitude.
It is convenient to absorb $\Delta,\, \epsilon$ and $\omega_c$ into a single
dimensionless parameter,
\begin{equation}\label{dimless1}
x_{\rm S}^{} = \left(\frac{\epsilon}{\omega_{\rm c}}\right)^K
\,\frac{ \Delta_{\rm S}}{\epsilon} \;. 
\end{equation}
The full rate $\gamma_n^\pm$ can now be written as
\begin{eqnarray}\label{series2}
\gamma_n^\pm &=& {\displaystyle \sum_{\ell =n}^{\infty} \gamma_{n,\ell}^\pm 
=\epsilon\, \sum_{\ell=n}^\infty  x_{\rm S}^{2\ell}
\,U_{n,\ell}^\pm \; ,} \\   \label{useries}
U_{n,\ell}^\pm  &=& {\displaystyle  \frac{1}{2^{2\ell}_{}}
\! \sum_{\{\xi_j\}'}\sum_{\{\eta_i\}'}\left[\cos\phi_\ell\,{\cal C}_\ell^{}
\pm \sin \phi_\ell\, {\cal S}_\ell^{} \right] }  \; .
\end{eqnarray}
The prime $\{\cdots\}'$ is to remind us that the double sum 
underlies the constraints (\ref{constraint2}). There are
$  \frac{(2\ell)!}{\ell! \,\ell!}
          \,       \frac{(2\ell)!}{(\ell + n)!(\ell - n)!}$
different charge sequences contributing to $U_{n,\ell}^\pm$.
All quantum fluctuations are in the $2\ell-1$-fold irreducible noise integrals
${\cal J}_{\ell}^{\pm}(\vec{\xi}\,)= {\cal C}_\ell^{}(\vec{\xi}\,) 
\pm i {\cal S}_\ell^{}(\vec{\xi}\,)$,
\begin{equation}\label{noiseint}
{\cal J}_\ell^{\pm}(\vec{\xi}\,) =
\int_0^\infty  \!\!\!\!\! {\rm d}^{2\ell-1}\vec{\tau}\,\,G_\ell^{({\rm c})}
(\vec{\tau},\vec{\xi}\,)\,e_{}^{\pm i \varphi_\ell^{}(\vec{\tau},\vec{\xi}\,)}\; ,
\end{equation}
where 
$  \int_0^\infty {\rm d}^{2\ell-1}\vec{\tau}\ldots   \equiv \int_0^\infty
{\rm d}\tau_{2\ell -1}^{}\cdots\, {\rm d}\tau_1^{}\,\ldots\, $. In terms of the
cumulative $p$ charges, the bias phase (\ref{biasphase}) reads 
\begin{equation}\label{phaserel}
\varphi_\ell^{}(\vec{\tau},\vec{\xi}\,) = \sum_{j=1}^{2\ell-1}p_j^{} \tau_j^{}  \;.
\end{equation}
The integrand of the noise integral ${\cal C}_\ell^{}$ (${\cal S}_\ell^{}$) depends on the
cosine (sine) of the bias phase 
and the interactions of the ``fluctuation'' charges are in the factor
$G_\ell^{(c)}(\vec{\tau},\vec{\xi}\,)$. With the subtractions in Eq.~(\ref{gsubtrac}),
all integrals in Eq.~(\ref{noiseint}) are convergent at large intervals between the charges.

The functions $U_{n,\ell}^{\pm}$ are linear combinations of 
the noise integrals ${\cal C}_{\ell}^{}(\vec{\xi}\,)$ and 
${\cal S}_{\ell}^{}(\vec{\xi}\,)$. As a consequence, the multitude of detailed 
balance relations (\ref{partdetbal}) for partial rates directly results in
a wealth of relations between noise integrals ${\cal J}_{\ell}^{\pm}(\vec{\xi}\,)$
of the same order. The coefficients of these relations include all the
friction effects. We conclude this section with the remark that it is only
the Ohmic scaling limit in which complete separation of friction and
quantum or classical noise takes place.

\section{Incoherent tunneling in the spin-boson model}\label{secinc}

The other quantum impurity model of interest is the one-channel anisotropic Kondo 
model~\cite{tsvelik} of spin $\frac{1}{2}$, which  describes scattering of fermions at an impurity 
with independent exchange constants $\rho J_\parallel^{}$ and $\rho J_\perp^{}$.
Here $\rho J_\parallel^{}$ conserves the polarization of the spin, while $\rho J_\perp^{}$ causes
spin-flip scattering. The AKM can also be expressed in bosonized form,
$H = H_0^{} + H_{\rm spin}^{}$, where $H_0^{}$ is given in Eq.~(\ref{hambsg}). The
boundary Hamiltonian is
\begin{equation}\label{hamspin}
H_{\rm spin}^{} = - \frac{\Delta_{\rm SB}^{}}{2}\,\left[\,\sigma_{+}^{}\, e^{i\phi(0)}_{}
\,+ \,\sigma_{-}^{}\,e^{-i\phi(0)}_{}\,\right] -\frac{\epsilon}{2} \sigma_z^{} \; .
\end{equation}
The additional bias term describes a local magnetic field energy. In general,
the relations between the parameters $g$ and $\Delta_{\rm SB}^{}$ of the boundary spin model
with the parameters $\rho J_\parallel^{}$ and $\rho J_{\perp}^{}$ of the AKM depend on the 
particular regularization prescription chosen for the singular scattering 
potential.~ \cite{weissbook,leggett87,tsvelik}

On the other hand, the boundary spin model (\ref{hamspin}) with (\ref{hambsg}) is the
polaron-transformed equivalent of the 
spin-boson (SB) model in the Ohmic scaling limit,~\cite{weissbook,leggett87}
\begin{equation}\label{tssham}
H_{\rm SB}^{} =  - \frac{ \Delta_{\rm SB}^{}}{2}\,\sigma_x^{}
-\frac{\epsilon}{2} \,\sigma_z^{} 
-\frac{1}{2} \sigma_z^{} \sum_{\alpha} c_\alpha^{}x_{\alpha}^{} + H_{\rm bath}^{} \;.
\end{equation}
In view of application of the SB model to a damped particle in a double well 
in the two-state limit, we choose  
the basis to be  formed by the localized states $|{\rm R}\rangle$ (right) 
and $|{\rm L}\rangle$  (left), which are eigenstates of the Pauli matrix
$\sigma_z^{}$ with eigenvalues 
$+1$ and $-1$. The potential drop from $| L\rangle$ to $|R\rangle$ is $\epsilon$, and
$\Delta_{\rm SB}^{}$ is the tunneling coupling.
Again, we have the correspondence $K=1/g$ and $\epsilon =2 \pi V$.

A valid dynamical quantity of interest is the expectation value of $\sigma_z$ at time
$t$ describing the difference of the populations of the two states,
$\langle\sigma_z(t)\rangle = P_{\rm R}(t) - P_{\rm L}(t)$. The dynamics of
$\langle\sigma_z(t)\rangle$ may be described in terms of the exact general master
equation (GME)
\[
\frac{{\rm d}\langle\sigma_z(t)\rangle}{{\rm d}t} = 
- \!\!\int_0^t\!\! {\rm d}t'\!\left[ K^{\rm (a)}(t-t') + K^{\rm (s)}(t-t')\langle
\sigma_z(t') \rangle \right] \; ,
\]
or equivalently in Laplace space
\begin{equation}\label{gme}
\lambda \langle\hat{\sigma}_z(\lambda)\rangle \, = \, - \hat{K}^{\rm (a)}(\lambda)
- \hat{K}^{\rm (s)}(\lambda) \langle\hat{\sigma}_z(\lambda)\rangle \; ,
\end{equation}
where the kernels $K^{\rm (s/a)}(\tau)$ and $\hat{K}^{\rm (s/a)}(\lambda)$ are 
even/odd under bias inversion.

In this section we are interested in the regime where the kernels decay fast
compared to the relevant time scales of the relaxation dynamics of the particle.
On this assumption, the GME becomes local in time,
\begin{equation}\label{master2}
\frac{{\rm d}\langle\sigma_z(t)\rangle}{{\rm d}t} =
\tilde{\gamma}^{-} -\, \tilde{\gamma}^{+} - \, (\tilde{\gamma}^{+} 
+ \tilde{\gamma}^{-}) \langle\sigma_z(t)\rangle \; ,
\end{equation}
where 
\begin{equation}
\begin{array}{rcl}
\tilde{\gamma}^\pm &=& {\displaystyle \frac{1}{2}\, \int_0^\infty {\rm d}\tau\,
\left[\,K^{\rm (s)}(\tau) \pm K^{\rm (a)}(\tau)\,\right] }  \\[4mm]
&=& {\displaystyle 
\frac{1}{2}\left[ \hat{K}^{\rm (s)}(\lambda=0) \pm \hat{K}^{\rm (a)}(\lambda=0)
\right]          }
\end{array}
\end{equation}
is the forward/backward incoherent tunnneling rate. 
The solution of Eq.~(\ref{master2}) is
\begin{equation}\label{solu1}
\langle\sigma_z(t)\rangle = \langle\sigma_z\rangle_{\rm eq}^{}
+\left[\langle\sigma_z (0)\rangle - \langle\sigma_z\rangle_{\rm eq}^{}\right]\,
e^{-(\tilde{\gamma}^{+} + \tilde{\gamma}^{-})\,t}_{}  \; ,
\end{equation}
where $\langle\sigma_z^{}\rangle_{\rm eq}^{}$ is the equilibrium value,
\begin{equation}
\langle\sigma_z\rangle_{\rm eq}^{} = \frac{\tilde{\gamma}^+ -\tilde\gamma^-}{
\tilde{\gamma}^{+} + \tilde{\gamma}^{-}} = \tanh\left(\frac{\epsilon}{2T}\right) \;.
\end{equation}
The second form follows from detailed balance.

The expression (\ref{solu1}) follows again from a cluster expansion for
$\langle\hat{\sigma}_z(\lambda)\rangle$. The obvious difference between the Schmid 
and the SB model is that in the former the time-ordered $u$ and $v$ charges may 
have arbitrary charge order, while in the latter they must alternate in sign.
In the SB model, the $2\ell + 1 $ $u$ charges and $2m + 1$ $v$ charges in (\ref{clus}) 
are therefore separately ordered as 
\begin{equation}\label{clus1}
{\cal A}_{\ell,m}^{\pm 1} \equiv \{\pm,\mp,\pm,\cdots,\pm;\; \pm,\mp,\pm,\cdots,\pm\}
\; .
\end{equation}
Thus, the series for the total tunneling rate in $\{u,\,v\}$ representation is
\begin{equation}
\tilde{\gamma}^\pm = \sum_{\ell =0}^\infty \sum_{m=0}^\infty
\gamma_1^\pm [{\cal A}_{\ell,m}^{\pm 1}]\;.
\end{equation}
The rate contribution $\gamma_1^{\pm}[{\cal A}_{\ell,m}^{\pm 1}]$ is given in
(\ref{partial1}), where $\Delta_{\rm SB}^{}$ takes the place of $\Delta_{\rm S}^{}$.
Thus, the full rate in the SB model corresponds to a particular partial rate
contributing to the rate $\gamma_1^{\pm}$ in the Schmid model.

Next, we switch to the $\{\eta,\xi\}$ representation.
For the set of paths specified by (\ref{clus1}) the sequence of $\eta$'s and $\xi$'s 
are restricted such that the cumulative charges $r_j^{}$ and $p_j^{}$ 
$(j= 1,\cdots,\,2+ 2\ell +2m)$
obey the constraints
\begin{equation}\label{etaxiconstr}
\begin{array}{rclcl}
0  &\le& r_j^{} \,\le\, 2 \;, \quad\qquad &\mbox{for} & \qquad {\cal A}_{\ell,m}^+  \; ,
\\[2mm]   
0  &\ge& r_j^{} \,\ge\, -2 \;,   &\mbox{for} &  \qquad  {\cal A}_{\ell,m}^-\;,  \\[2mm]  
p_j^{} &=& 0\; ,        & \mbox{for}&    \qquad j\;\mbox{even}\; .
\end{array}
\end{equation}
In analogy with Eq.~(\ref{dimless1}) we introduce the dimensionless dressed
tunneling coupling as
\begin{equation}\label{dimless2}
x_{\rm SB}^{} = \left(\frac{\epsilon}{\omega_c}\right)^K \,
\frac{\Delta_{\rm SB}}{\epsilon} \; .
\end{equation}
Formally we may then write
\begin{equation}\label{wtexpan}
\tilde{\gamma}^\pm = \sum_{\ell=1}^\infty \tilde{\gamma}_\ell^\pm \; ; \qquad 
\tilde{\gamma}_\ell^\pm = \epsilon\,x_{\rm SB}^{2\ell}\, W_{\ell}^\pm \; ,
\end{equation}
where $W_{\ell}^\pm = U_{1,\ell}^\pm\,$, and where the double sum in 
(\ref{useries}) underlies the SB constraints (\ref{etaxiconstr}). The $\eta$ sum 
in $U_{1,\ell}^{\pm}$ is easily performed, yielding
\begin{equation}\label{wellexpr}
W_\ell^\pm = {\textstyle\left(-\frac{1}{2}\right)^{\ell-1} } \!
\cos^{\ell}(\pi K) \!\! \sum_{ \{\xi_j\}''} \!\!
\left[ {\cal C}_{\ell}^{}  \pm \xi_1 \tan(\pi K) {\cal S}_{\ell}^{} \right] \, ,
\end{equation}
where $\{ \cdot\}''$ reminds us of the SB constraint
$\xi_{2k}^{} = - \xi_{2k-1}^{}$ $(k=1,\cdots,\,\ell)$ [\,cf. Eq.~(\ref{etaxiconstr})\,].
 
\section{Weak-tunneling expansion at $T=0$}\label{secwt}

In this section we investigate the weak-tunneling expansion for the
tunneling rates in the Schmid and SB model. 
We restrict the attention to the scaling limit and zero temperature. Then the
charge interaction is logarithmic at all distances, $Q(\tau) = Q_0^{}(\tau)$ where 
\begin{equation}\label{inter3}
Q_0^{}(\tau)= 2K \ln(\tau) + i\,\pi K\,{\rm sgn}(\tau) \; .
\end{equation}
Consequently, the noise integrals ${\cal J}_\ell^{\pm}(\vec{\xi})$ given in
Eq.~(\ref{noiseint}) diverge at short 
distances between neighbouring attractive charge pairs for $K\ge \frac{1}{2}$.
Now we first focus the attention on the case $K< \frac{1}{2}$, in which all noise integrals 
are finite. The way to regularize the integrals for $K\ge \frac{1}{2}$ without introducing
nonuniversal cutoff-dependent physics shall be discussed at the end of
Subsection \ref{secbsg}.

At $T=0$, backward transitions to higher wells are absent, i.e.,
\begin{equation} \label{ratezero}
\gamma_n^{-}[\alpha_{\ell,m}^{-n}] =0 \qquad\mbox{for all}\qquad \alpha_{\ell,m}^{-n}
\; ,
\end{equation}
as follows from Eq.~(\ref{partdetbal}). In the sequel, the $u$ charges are depicted as
$+$ and $-$, and the $v$ charges as $\oplus$ and $\ominus$.
Complex conjugation is synomynous with exchange of the $u$ and $v$ charges, and
charge conjugation means change of the signs of the charges.

\subsection{Golden rule limit}\label{secgr}

For very weak intersite coupling $\Delta_{\rm S}^{}$, the leading contribution in the
series (\ref{series2}) has one $u$ and one $v$ charge. The respective ordered arrangements 
are $+\;\oplus$ and $\oplus\;+$ for the forward rate, and the
corresponding charge-conjugate pairs for the backward rate. The
noise integrals associated have $\xi_1 = p_1 = \pm 1$ and read
\begin{equation}\label{noisint}
J^\pm_{1} \equiv C_1^{} \pm i S_1^{} = \int_0^\infty {\rm d}\tau\, e^{\pm i \tau}_{} 
\frac{1}{\tau^{2K}_{}} \; .
\end{equation}
Equation~(\ref{series2}) with (\ref{useries}) for $n=\ell =1$ then gives
\begin{equation}\label{grlim1}
\gamma _{1,1}^\pm = \epsilon \frac{x_{\rm S}^2}{2} \left[\,\cos(\pi K)\, C_1^{} \pm
\sin(\pi K)\, S_1^{}\,\right] \; .
\end{equation}
With the substitution $\tau = \pm i x$ in Eq.~(\ref{noisint}), it is
straightforward  to see that
\begin{equation}\label{cs1}
C_1^{} = \tan(\pi K)\, S_1^{} \; ,
\end{equation}
and $S_1^{} = \cos(\pi K)\Gamma(1 - 2K)$. Thus $\gamma_{1,1}^{-} =0$, and
\begin{equation} \label{grschm}
\gamma_{1,1}^+ = \epsilon \,x_{\rm S}^2 \sin(\pi K)\, S_1^{}  
 =  \frac{\pi}{2\Gamma(2K)}\,\epsilon \, x_{\rm S}^2 \;. 
 \end{equation}
This is the golden rule tunneling rate to the nearest-neighbour site in forward direction
in the Schmid model.
Clearly, the same expression 
\begin{equation}\label{grtss}
\tilde{\gamma}_1^{+} = \epsilon\, x_{\rm SB}^2 \sin(\pi K) S_1^{} 
\end{equation}
holds for the rate in the SB model in the nonadiabatic limit.

\subsection{The order $\Delta^4_{}$}\label{secdeltafour}

In order $\Delta_{\rm S}^{4}$ in the Schmid model, there is (i) the possibility for direct 
tunneling transitions to the next-to-nearest-neighbour site, and (ii) the possibility 
for tunneling transitions to the nearest-neighbour site via a pair of 
virtual forward/backward moves. The first possibility contributes to the rate
$\gamma_2^{\pm}$, and the second one to the rate $\gamma_1^{\pm}$.
In case (i), the charge arrangements have two $u$ charges on the path $q$
and two $v$ charges on the path $q'$, all of the same sign.
The unrestricted integration over $\tau$ in Eq.~(\ref{partial1}) yields
all possibilities of mixing up the $u$ with the $v$ charges.
The resulting sets ${\cal D}_1^{}$ and $\bar{\cal D}_1^{}$ each have six
different time-ordered arrangements as shown in Fig.~\ref{fig1}. These sets make up the
backward rate $\gamma_{2,2}^{-}$ and forward rate $\gamma_{2,2}^{+}$, respectively. 
\begin{figure}[ht]
\[
\begin{array}[t]{ccccccccc}
  &   & {\cal D}_1^{}&    &             &   &  & \bar{\cal D}_1^{} &  \\[2mm]
- & - & \ominus & \ominus & \qquad\qquad \qquad & + & + & \oplus & \oplus \\[1mm]
- & \ominus & - & \ominus & \qquad\qquad\qquad & + & \oplus & + & \oplus \\[1mm]
- & \ominus & \ominus & - & \qquad\qquad \qquad & + & \oplus & \oplus & +  \\[1mm]
\ominus & - & - & \ominus & \qquad\qquad \qquad & \oplus & + & + & \oplus \\[1mm]
\ominus & - & \ominus & - & \qquad\qquad \qquad & \oplus   & + & \oplus & + \\[1mm]
\ominus & \ominus & - & - & \qquad\qquad \qquad & \oplus & \oplus & + & +  
\end{array}
\] 
\caption{\label{fig1}
The set ${\cal D}_1^{}$ consisting of six different charge sequences
or paths constitutes the backward rate $\gamma_{2,2}^{-}$. The charge-conjugate
set $\bar{\cal D}_1^{}$ determines the forward rate $\gamma_{2,2}^{+}$. }
\end{figure}

In case (ii), there are three charges on path $q$ and one charge on path $q'$, and
vice versa. Accordingly, there are three different sets denoted by 
${\cal D}_2,\,{\cal D}_3,\, {\cal D}_4$,
and the respective complex conjugate counterparts ${\cal D}_2^\ast,\,{\cal D}_3^\ast,\,
{\cal D}_4^\ast$. These six sets make up the rate $\gamma_{1,2}^{-}$ (see Fig.~\ref{fig2}).
\begin{figure}[ht]
\[
\begin{array}[t]{cccccccccccc}
  &   &{\cal D}_2^{} & & & & {\cal D}_3^{} & & & & {\cal D}_4^{} & \\[2mm] 
- & - & + & \ominus &\qquad\quad -&+&-&\ominus&\qquad\quad +&-&-&\ominus\\[1mm]
- & - & \ominus & + &\qquad\quad -&+&\ominus&-&\qquad\quad +&-&\ominus&-\\[1mm]
- & \ominus & - & + &\qquad\quad -&\ominus&+&-&\qquad\quad +&\ominus&-&-\\[1mm]
\ominus & - & - & + &\qquad\quad \ominus&-&+&-&\qquad\quad \ominus&+&-&-
\end{array}
\]
\caption{\label{fig2} The three sets ${\cal D}_2^{},\,{\cal D}_3^{},\,{\cal D}_4^{}$, 
each consisting of four different arrangements, form, together with their complex conjugate 
counterparts ${\cal D}_2^\ast\,{\cal D}_3^\ast,\,{\cal D}_4^\ast$, the 
rate $\gamma_{1,2}^{-}$.  }  
\end{figure}

The quantum fluctuations are carried by three different arrangements of the
cumulative charges $p_1^{},\,p_2^{},\,p_3^{}$, which are $\{1,\,2,\, 1\}$,
$\{1,\,0 ,\, 1 \}$, and $\{1,\,0 ,\, -1 \}$, and their charge conjugate 
counterparts. The corresponding connected complex noise integrals  
\begin{equation}
J_{2,k}^{\pm} = C_{2,k}^{} \pm i S_{2,k}^{}\;, \qquad k=1,\,2,\,3 
\end{equation}
are 
\begin{eqnarray}\nonumber
J_{2,1}^\pm \!\!   &=& \!\!\int_0^\infty \!\!\!\!\!{\rm d}^3 \!\vec{\tau}\,
e^{\pm i (\tau_1^{} +2\tau_2^{} +\tau_3^{})}_{}
\left( \frac{\tau_1^{} \tau_3^{}}{\tau_2^{} \tau_{12}^{} \tau_{23}^{} 
\tau_{13}^{}}\right)^{2K},  \\[1mm]              \nonumber
J_{2,2}^\pm  \!\! &=&\!\! \!\int_0^\infty \!\!\!\!{\rm d}^3\!\vec{\tau}\,
e^{\pm i(\tau_{1}^{} +\tau_{3}^{})}_{}
\!\left[\!
\left( \frac{\tau_{12}^{} \tau_{23}^{}}{\tau_1^{} \tau_{2}^{} \tau_{3}^{} 
\tau_{13}^{}}\right)^{2K} \!\!-\left( \frac{1}{\tau_1^{} \tau_3^{}}\right)^{2K} 
\right],    \\[1mm]    \nonumber
J_{2,3}^\pm \!\! &=& \!\!\! \int_0^\infty \!\!\!\!{\rm d}^3\!\vec{\tau}\,
e^{\pm i(\tau_1^{} - \tau_3^{})}_{}
\! \left[ \! \left( \frac{\tau_{2}^{} \tau_{13}^{}}{\tau_1^{} \tau_{3}^{} \tau_{12}^{} 
\tau_{23}^{}}\right)^{2K} \!\!- \left(\frac{1}{\tau_1^{} \tau_3^{}}\right)^{2K}\right].
\end{eqnarray}
The integrals $J_{2,2}^{\pm}$ and $J_{2,3}^{\pm}$ embody intermediate visits of 
diagonal states. The subtractions made serve to eliminate the reducible parts.
The set ${\cal D}_1^{}$ yields the backward rate $\gamma_{2,2}^{-}$, which is zero.
Upon transcribing the set ${\cal D}_1^{}$ into the $\{\eta,\xi\}$ representation, we get
the following linear relation between noise integrals $J_{2,k}^{\pm}$ ($k=1,\,2,\,3$),
\begin{equation}\label{lineq1}
\begin{array}{rcl}
 e^{-i 4\pi K}_{} J_{2,1}^{-} &+& e^{i 4\pi K}_{} J_{2,1}^{+} + 
e^{- i 2\pi K}_{} J_{2,2}^{-}  \\[3mm]  
&& \quad  + \;e^{i 2\pi K}_{}J_{2,2}^{+} + J_{2,3}^{-} + J_{2,3}^{+} = 0 \; .
\end{array}
\end{equation}
The sets ${\cal D}_2^{},\,{\cal D}_3^{},\,{\cal D}_4^{}$ yield complex partial ``rates''
$\gamma_{1,1}^{-}[{\cal D}_2^{}]$, $\gamma_{1,1}^{-}[{\cal D}_3^{}]$, and
$\gamma_{1,1}^{-}[{\cal D}_4^{}]$. According to Eq.~(\ref{partdetbal}), these partial rates 
vanish individually at $T=0$. The resulting linear relations between the noise 
integrals are
\begin{equation}\label{lineq2}
\begin{array}{rcl}
\left(e^{-i2\pi K}_{} + e^{-i 4\pi K}_{}\right)\! J_{2,1}^{-} 
+ e^{-i 2\pi K}_{} J_{2,2}^{-} 
+J_{2,3}^+ \!\! &=& \!\! 0 \, , \\[3mm] 
 e^{-i 2\pi K}_{} J_{2,2}^{-} + J_{2,2}^{+} 
+ \left(1+ e^{-i 2\pi K}_{}\right) J_{2,3}^- \!\! &=& \!\! 0 \, ,
\\[3mm]
\left( 1 + e^{i 2\pi K}_{}\right) \! J_{2,1}^{+} + J_{2,2}^{+} 
+ e^{-i 2\pi K}_{} J_{2,3}^{+} \!\! &=& \!\! 0 \, .
\end{array}
\end{equation}

In addition, there are relations between the real and imaginary parts of the noise integrals
$J_{2,1}^\pm $ and $J_{2,2}^\pm$.  Upon rotating the integration paths by 
$\pm \frac{\pi}{2}$,  $\vec{\tau}= \pm i\vec{x}$, the integrals
$J_{2,1}^\pm $ and $J_{2,2}^\pm$ receive a definite phase factor $\mp i e^{\mp i 2\pi K}_{}$.
Thus we get
\begin{equation}\label{lineq3}
\begin{array}{rcl}
 C_{2,1}^{} &=& \tan(2\pi K)\, S_{2,1}^{}\;,  \\[2mm]      
C_{2,2}^{} &=& \tan(2\pi K) \, S_{2,2}^{} \; . 
\end{array}
\end{equation}
Furthermore, for reasons of symmetry we have
\begin{equation}
S_{2,3}^{} = 0 \; . \label{lineq4}
\end{equation}

The set of linear equations (\ref{lineq1}) and (\ref{lineq2}), together with
expressions (\ref{lineq3}) and (\ref{lineq4}), yields the relations
\begin{equation}\label{resord4}
\begin{array}{rcl} 
C_{2,3}^{} &=& \tan(2\pi K)\, S_{2,1}^{} \; ,  \\[2mm]   
S_{2,2}^{} &=& - [\,1 + \cos(2\pi K)\,]\,S_{2,1}^{} \;.
\end{array}
\end{equation}

With the expressions (\ref{lineq3}) - (\ref{resord4}) the noise integrals 
$C_{2,1}^{},\,C_{2,2}^{},\,C_{2,3}^{}$ and $S_{2,2}^{}$ can be expressed in terms 
of the noise integral $S_{2,1}^{}$. As a result, also the individual partial rates in forward 
direction can be written in terms of the noise integral $S_{2,1}^{}$. 
The nearest-neighbour partial ``rates'' in order $\Delta_{\rm S}^{4}$ are found to read
\begin{eqnarray*}
\gamma_{1,2}^{+} [\,\bar{\cal D}_2^{} + \bar{\cal D}_2^\ast\,]
 &=& - \epsilon \frac{x_{\rm S}^4}{4}\sin(2\pi K)\cos(2\pi K)\,S_{2,1}^{}\; , \\
\gamma_{1,2}^{+} [\,\bar{\cal D}_3^{} + \bar{\cal D}_3^\ast\,]
 &=& \epsilon \frac{x_{\rm S}^4}{4}\sin(2\pi K)[\, 1 + \cos(2\pi K)\,]\,S_{2,1}^{}\; , \\
\gamma_{1,2}^{+} [\,\bar{\cal D}_4^{} + \bar{\cal D}_4^\ast\,]
 &=& - \epsilon \frac{x_{\rm S}^4}{4}\sin(2\pi K) \,S_{2,1}^{} \; .
\end{eqnarray*}
Now observe that these partial rates add up to zero,
\begin{equation}
\gamma_{1,2}^{+} = 0 \;.
\end{equation}

Next, we recall that the set $\bar{\cal D}_1^{}$ in Fig.~\ref{fig1} determines the forward 
rate to the next-to-nearest-neighbour site in order $\Delta_{\rm S}^{4}$,
$\gamma_{2,2}^+$, in the Schmid model. Using again  the relations 
(\ref{lineq3}) - (\ref{resord4}), we obtain the compact expression
\begin{equation}\label{rschmid2}
 \gamma_{2,2}^{+} = - \epsilon \frac{x_{\rm S}^4}{2} \sin^2(\pi K)\sin(2\pi K) 
S_{2,1}^{}\; .
\end{equation}

Finally, we turn to the SB model. Because the signs of the $u$ and $v$ charges must 
alternate, the only
charge arrangement contributing to the rate $\tilde{\gamma}_2^{+}$ are the sets
$\bar{\cal D}_3^{}$ and $\bar{\cal D}_3^\ast$ (cf. Fig.~\ref{fig2}). Thus we obtain
\begin{equation}\label{rtss2}
\tilde{\gamma}_2^+ = \epsilon \frac{x_{\rm SB}^4}{4}\sin(2\pi K)
[\,1 + \cos(2\pi K)\,]\, S_{2,1}^{} \;. 
\end{equation}
Upon combining the expression (\ref{rschmid2}) with (\ref{rtss2}) we find the
simple relation
\begin{equation}\label{ratio2}
\tilde{\gamma}_2^+ = - \left(\frac{\Delta_{\rm SB}^{}}{2\sin(\pi K) \Delta_{\rm S}^{}}
\right)^4\, 4\sin^{2}_{}(2\pi K)\,\gamma_{2,2}^{+} \; .
\end{equation}
This concludes the discussion of order $\Delta^4_{}$.

\subsection{The order $\Delta^6$}\label{secdeltasix}

In the Schmid model, the irreducible clusters of order $\Delta^6$ contribute to the 
rates $\gamma_3^\pm,\, \gamma_2^\pm$ and $\gamma_1^\pm$. Now consider first the
various noise integrals of this order.
There are 10 different complex conjugate pairs of possibilities to order the 
$\xi$ charges. Some of these pairs can be transferred into each other by
reversal of the order of the charges. As a result, there are seven different
arrangements left. The related noise integrals
$J_{3,k}^\pm = C_{3,k}^{} \pm i S_{3,k}^{}$ $(k=1,\,\cdots,\, 7)$ are listed in 
Fig.~\ref{fig3}. The individual expressions can be be traced from
Eq.~(\ref{noiseint}). The noise integrals $J_{3,1}^\pm$ and $J_{3,2}^\pm$ are irreducible 
on principle, and we assume that the reducible parts of 
$J_{3,3}^\pm,\,\cdots,\, J_{3,7}^\pm$ are already subtracted.

\begin{figure}[ht]
\[
\begin{array}[t]{rrrrrrrr}
p_1^{} &\quad p_2^{} & \quad p_3^{} &\quad p_4^{} & \quad p_5^{}    \\[2mm]
1      &        2        &    3   &    2  &   1   & \quad\qquad J_{3,1}^\pm   \\[1mm]
1      &        2        &    1   &    2  &   1   & \quad\qquad J_{3,2}^\pm   \\[1mm]
1      &        2        &    1   &    0  &   1   & \quad\qquad J_{3,3}^\pm   \\[1mm]
1      &        0        &    1   &    0  &   1   & \quad\qquad J_{3,4}^\pm   \\[1mm]
1      &        2        &    1   &    0  &  -1   & \quad\qquad J_{3,5}^\pm   \\[1mm]
1      &        0        &   -1   &    0  &   1   & \quad\qquad J_{3,6}^\pm   \\[1mm]
1      &        0        &    1   &    0  &  -1   & \quad\qquad J_{3,7}^\pm       
\end{array}
\]
\caption{\label{fig3}The different sets of cumulative charges, which (together
with the charge-conjugate counterparts) define the noise integrals
$J_{3,1}^\pm,\,\cdots,\,J_{3,7}^\pm$. }
\end{figure}

The cumulative charges of the noise integrals $J_{3,1}^\pm,\cdots,\,J_{3,4}^\pm$ do not 
change sign. The real parts of these integrals are again directly related to the 
respective imaginary parts. The corresponding relations are discovered upon rotating the
integration paths by $\pm \frac{\pi}{2}$, $\vec{\tau} = \pm i \vec{x}$. The integrals then 
receive  a definite phase factor $\pm i e^{\mp i 3\pi K}_{}$. We then find
\begin{equation}\label{cs3rel}
C_{3,k}^{} = \tan(3 \pi K) S_{3,k}^{}\;, \quad \mbox{for}\quad k= 1,\,2,\,3,\, 4 \;.
\end{equation}

Now consider first the rate contribution $\gamma_{3,3}^{\pm}$. The relevant charge
sequences have three charges of type $u$ and three charges of type $v$, all of the
same sign. There are 20 different possibilities to mix up the $v$ charges with the
$u$ charges, i.e., there are 20 individual paths on the $\{q,q' \}$ or $\{\eta, \xi\}$
plane. Hence $\gamma_{3,3}^{\pm}$ emerges as a linear combination of
20 noise integrals chosen from the list displayed in Fig.~\ref{fig3},
and the coefficients can be traced from Eq.~(\ref{useries}).

Secondly, the rate $\gamma_{2,3}^\pm$ is made up of arrangements with 
four $u$ charges (three of the same and one of opposite sign) and two $v$ charges of 
same sign, and of the corresponding arrangements in which the $u$ and $v$ charges 
are interchanged.
According to the four different possibilities to order the four $u$ charges, there are 
four different complex partial ``rates'', denoted by $\gamma_{2,3}^\pm[k]$ 
$(k=1,\cdots,\,4)$.
There are 15 different possibilities to mix up the four $u$ charges with the two $v$ charges.
Hence each of the partial rates $\gamma_{2,3}^\pm[k]$ can be written as a linear 
combination of 15 noise integrals taken from the set $J_{3,1}^\pm,\,\cdots,\,J_{3,7}^\pm$.
Taking into account also the complex conjugate arrangements, the rate $\gamma_{2,3}^\pm$ is
made up by altogether 120 different charge sequences or tunneling paths.

Thirdly, the rate $\gamma_{1,3}^\pm$ emerges from (i) the sets with five $u$ charges
(three of the one and two of the other sign),
and one $v$ charge, (ii) the corresponding complex conjugate sets, 
and (iii) the sets with three $u$ and three $v$ charges 
(each with two charges of the one sign and one charge of the other sign).
In case (i) there are 10 possibilities to order the $u$ charges. This results in
10 complex partial rates $\gamma_{1,3}^\pm[k]$, $k =1,\,\cdots,\, 10$. Each 
is a linear combination of six noise integrals, according to the six possibilities
to add in the $v$ charge to the arrangement of $u$ charges. Case (ii) yields the
respective complex conjugate counterparts. Case (iii) gives three possibilities to 
order the $u$ and three possibilities to order the $v$ charges.
Accordingly, there are additional 9 partial rates 
$\gamma_{1,3}^\pm[k]$, $k=11,\,\cdots,\,19$. Since there are 20 different possibilities 
to mix up the $u$ with the $v$ charges, each of these partial rates is a linear combination 
of 20 noise integrals, again taken 
from the list shown in Fig.~\ref{fig3}. In summary, there are 300 different charge sequences
or tunneling paths altogether which determine the partial rate $\gamma_{1,3}^\pm$.

All the noise integrals $J_{3,j}^\pm$ $(j=1,\,\cdots,\, 7)$ are related with each 
other such that
\begin{eqnarray}\nonumber 
\gamma_{3,3}^{-} &=& 0 \; ,  \\   \label{linrel}
\gamma_{2,3}^{-}[k] &=& 0 \, ,\,\quad\qquad k=1,\,\cdots,\, 4 \; , \\  \nonumber
\gamma_{1,3}^{-}[k] &=& 0 \,, \quad\qquad k=1,\,\cdots,\,19 \; .
\end{eqnarray}

With use of Eqs.~(\ref{linrel}) and (\ref{cs3rel})
the noise integrals $J_{3,3}^\pm,\, \cdots,\,J_{3,7}^\pm$ can be expressed in 
terms of the noise integrals $S_{3,1}^{}$ and $S_{3,2}^{}$. Upon using these relations
the respective forward rates are found as
\begin{equation}\label{resorder31}
\begin{array}{rcl}
\gamma_{1,3}^+ &=& 0 \;, \\[3mm] 
\gamma_{2,3}^+ &=& 0 \; , \\[2mm]
\gamma_{3,3}^+ &=& {\displaystyle
\epsilon \frac{x_{\rm S}^6}{3} \sin^2(\pi K)\sin^2(2\pi K)} \\[4mm]
&&\times\;[\,\sin(\pi K)S_{3,2}^{} + \sin(3\pi K)S_{3,1}^{}\,] \; .  
\end{array}
\end{equation}

Finally, consider the forward rate in the SB model in order $\Delta^6$. 
Since the $u$ and $v$ charges must alternate in sign separately,
there are only 32 different tunneling paths 
contributing to $\tilde{\gamma}_3^{+}$, and the noise integrals involved are
$J_{3,4}^\pm$, $J_{3,6}^{\pm}$ and $J_{3,7}^{\pm}$. Expressing these in
terms of the noise integrals $S_{3,1}^{}$ and $S_{3,2}^{}$ and using 
Eq.~(\ref{resorder31}), we  find
\begin{equation}\label{tssrate3}
\tilde{\gamma}_3^{+} = \left(\frac{\Delta_{\rm SB}}{2\sin(\pi K)\Delta_{\rm S}}
\right)^6 4 \sin^2(3\pi K) \gamma_{3,3}^{+} \;.
\end{equation}

In conclusion, we have found again formidable cancellations
between the contributions of the various tunneling paths in the Schmid model.
All those rates of order $\Delta_{\rm S}^6$ which 
involve pairs of virtual forward/backward moves vanish. 
In addition, we found again a simple relation
between the rate in the SB model and the one nonvanishing rate in the Schmid model
of same order in $\Delta$. 

\subsection{Higher orders}\label{sechigherord}

With increasing order of the perturbative expansion, the number of contributing
tunneling paths increases enormously.
Therefore, the computation, although straightforward, becomes
rather elongated. The general strategy is based on two observations:

(i) The set of equations (\ref{ratezero}) results in a set of
linear relations between the noise integrals ${\cal J}_\ell^{\pm}(\vec{\xi}\,)$ 
of same order $\ell$ and different $\vec{\xi}$.

(ii) Consider an arrangement of ``fluctuation'' charges 
$\vec{\xi}=\vec{\xi}^{\rm \,(d)}_{}$ for which the bias phase
$\varphi_\ell^{}$ has a definite sign.
We see from Eq.~(\ref{phaserel}) that the sign of $\varphi_\ell^{}$
is definite when all cumulative charges $p_j^{}$ have the same sign (apart from zeros),
say ${\rm sgn}(p_j^{})=+$.
It is now decisive that the real and imaginary part of each individual noise integral 
${\cal J}_\ell^\pm(\vec{\xi}^{\rm \,(d)}_{}\,)$ are related to each other. 
To determine the relation, we rotate the integration paths
by $\pm \frac{\pi}{2}$, $\vec{\tau} = \pm i \vec{x}$. 
With these rotations,
${\cal J}_\ell^\pm(\vec{\xi}^{\rm \,(d)}_{}\,)$ picks up
a definite phase factor, which is  $\mp i\,e^{\pm i\pi(1-K)\ell}_{}$. Thus we find
\begin{equation}\label{csgenrel}
{\cal C}_{\ell}^{}(\vec{\xi}^{\rm \,(d)}_{}\,) = 
\tan (\ell \pi K )\,{\cal S}_{\ell}^{}(\vec{\xi}^{\rm \,(d)}_{}\,) \; .
\end{equation}
This is the generalization of Eq.~(\ref{cs3rel}) to order $\Delta^{2\ell}_{}$.

Property (i) is in general a virtue of the Ohmic scaling limit, Eq.~(\ref{inter2}).
It also holds at finite temperatures, where Eq.~(\ref{ratezero}) is replaced by
Eq.~(\ref{partdetbal}) (see remark at the end of Section \ref{sec3}).
Property (ii) is based on the scale invariant logarithmic charge 
interaction (\ref{inter3}). It does not hold when scale 
invariance is broken by finite temperature.

Upon combining (i) with (ii) we find for fixed $\ell$ that all those noise integrals
${\cal J}_\ell^{\pm}(\vec{\xi}\,)$, which have one or several of the cumulative charges
$p_j^{}$ equal to zero, can be expressed in terms of noise integrals 
${\cal S}_\ell^{}(\vec{\xi}\,)$ in which all $p_j^{}$ are nonzero, which entails that the
respective $p_j^{}$ have all the same sign. Making use of these
relations, we discover formidable cancellations among the various path contributions
to $U_{n,\ell}^{+}$ in Eq.~(\ref{useries}). In fact, all tunneling
paths which describe detours via virtual forward/backward hops cancel each other.
Thus we find
\begin{equation}\label{gamplusnull}
\gamma_{n,\ell}^{+} = 0\;, \qquad\quad\mbox{for}\qquad\ell > n \;.
\end{equation}
Only direct tunneling paths which have all $u$ and $v$ charges positive contribute.
Finally, we obtain in generalization of Eq.~(\ref{resorder31})
\begin{equation}\label{res1}
\begin{array}{rcl}
\gamma_n^{+} &=& \gamma_{n,n}^{+}= \epsilon\,x_{\rm S}^{2n} U_{n,n}^{+}\;,\\[3mm]
U_{n,n}^{+} &=& {\displaystyle \frac{(-1)^{n-1}}{2 n} 
\sum_{\{\xi_j\}'} {\cal S}_n^{}(\vec{\xi}\,)
\prod_{k=1}^{2n-1} \sin(\pi K p_k^{})  }      \; .
\end{array}
\end{equation}

Next we turn to the discussion of the SB model. Making again use of the 
relations between the noise integrals of same order $n$, we discover that the weight
$W_n^{+}$ of the SB partial rate $\tilde{\gamma}_n^+ = \epsilon x_{\rm SB}^{2n}W_n^+$ 
 given in Eq.~(\ref{wellexpr}) can directly be expressed
in terms of the weight $U_{n,n}^{+}$ of the rate 
$\gamma_n^+ = \epsilon\,x_{\rm S}^{2n} U_{n,n}^{+}$ in the Schmid model. We find
\begin{equation}\label{uwrel}
W_n^+ = (-1)^{n-1} \frac{4\sin^2(n\pi K)}{[2\sin (\pi K)]^{2n}}\, 
U_{n,n}^+ \; .
\end{equation}                   

For later convenience, we now introduce a frequency scale $\epsilon_0^{}$ analogous 
to the Kondo scale in the Kondo model and the scale $T_{\rm B}'$ in 
QIPs.~\cite{fendley95b}
The relation of $\epsilon_0^{}$ to the bare parameters of the Schmid model is
established by the exact self-duality of this model.\cite{weissbook}
As far as the SB is concerned, it is convenient to absorb the factor
$[2\sin(\pi K)]^{-2n}$ occurring in Eq.~(\ref{uwrel}) into a corresponding redefinition
of the Kondo scale. Thus we have
\begin{equation}\label{kondoscale}
\epsilon_0^{2-2K} = \frac{2^{2-2K}_{} \pi^2}{\Gamma^2(K)}\,
\frac{\Delta_{\rm S}^2}{ 
\omega_{\rm c}^{2K} } = \frac{\Gamma^2(1-K)}{2^{2K}_{}} \frac{ \Delta_{\rm SB}^2}{ 
\omega_{\rm c}^{2K}}    \; .
\end{equation}
At fixed renormalized coupling $\epsilon_0^{}$, the SB
bare coupling $\Delta_{\rm SB}^{}$ is then related
to the bare coupling $\Delta_{\rm S}^{}$ of the Schmid model as 
\begin{equation}\label{deltarel}
\frac{\Delta_{\rm SB}^{}}{ \Delta_{\rm S}^{}}\, =\, 
\frac{ x_{\rm SB}^{}}{x_{\rm S}^{}}\, = \, 2 \sin(\pi K) \; .
\end{equation} 

Upon employing relation (\ref{uwrel}), we then find that the partial rate 
$\tilde{\gamma}_n^+$ of the SB model is simply given by
\begin{equation}\label{raterel}
\tilde{\gamma}_n^+ = (-1)^{n-1} 4\sin^2(n\pi K)  \, \gamma_n^+   \; .
\end{equation}
The intriguing functional expression (\ref{raterel}) is a major result of this paper.
It relates the incoherent partial tunneling rate
of order $\Delta_{\rm SB}^{2n}$ in the SB model to the full tunneling rate 
from site 0 to site $n$ in the Schmid model.

At the end of this subsection, let us briefly pause to shed light on the relations
(\ref{kondoscale}) and (\ref{deltarel}) from different perspective. First, we note that
the relation (\ref{deltarel}) is in agreement with findings from a unification of
the SB and Schmid model within the quantum group $SU(2)_q$.\cite{fendley95a,fendley95c}
Secondly, we recall the relationship  of the Ohmic SB model with the Kondo model.
While the original isotropic Kondo model is at $K=1$, the anisotropic Kondo model with
independent exchange constants $\rho J_\parallel$ and $\rho J_\perp^{}$ describes 
variable $K$. The correspondence between the two models is universal, i.e. it is
independent of the regularization prescription chosen for the singular
scattering potential,\cite{tsvelik} in the regime
$\rho J_\perp^{} =\Delta_{\rm SB}^{}/\omega_c^{}\ll 1 $ and
$\rho |J_\parallel| = |1-K| \ll 1$. In the parameter range 
$\rho J_\perp^{} \ll \rho J_\parallel \ll 1$, the Kondo energy (modulo a factor $2/\pi$)
is expressed in terms of the bare parameters as\cite{tsvelik,weissbook}
\begin{equation}\label{kondouniv}
\epsilon_0^{} = \left(\frac{\rho J_{\perp}^{}}{2\rho J_{\parallel}}
\right)^{1/\rho J_\parallel} \!\!\omega_c^{}\; =\; \left(
\frac{1}{(1-K)}\frac{ \Delta_{\rm SB}^{}}{(2\omega_c)^{K}_{}}\right)^{\frac{1}{1-K}} \; .
\end{equation}
The second form ensues from use of the correspondence relations. Now observe that
the second form in Eq.~(\ref{kondoscale}) coincides with the expression (\ref{kondouniv})
in the regime around the critical coupling, $|1-K| \ll 1$. These two independent conclusions
strongly confirm the above analysis of the perturbative expansions of the Schmid and SB
model.

\subsection{Statistical fluctuations}\label{secfluc}

From the characteristic function (\ref{charfunc1}) at $T=0$ we see 
that the connected moments of the probability current 
$\langle I\rangle_c^{} =  \sum_n I_n^+ = \sum_n n \gamma_n^+$ take the form
\begin{equation}
\langle I^{(m)}_{}\rangle_c = \sum_{n=1}^\infty n^{m-1}_{} I_n^{+} \; ,
\end{equation}
where what is called $\langle I^{(m)}_{}\rangle_c$ corresponds to
$\langle n^{m}_{}(t)\rangle_c/t$.
Now, since $I_n^{+}/\epsilon$ is of order $\Delta_{\rm S}^{2n}\epsilon_{}^{(2K-2)n}$,
as follows from Eq.~(\ref{res1}) with Eq.~(\ref{dimless1}), simple moment relations can be 
derived. We find
\begin{equation}\label{curmom}
\begin{array}{rcl}
\langle I_{}^{(m)} \rangle_c &=& {\displaystyle
\left(\frac{\Delta_{\rm S}^{}}{2}\frac{\partial}{\partial
\Delta_{\rm S}^{}}\right)^{m-1} \langle I \rangle_c   }  \\[3mm]
&=& {\displaystyle
\epsilon \left( \frac{\epsilon}{2(K-1)}\, \frac{\partial}{\partial\epsilon}\right)^{m-1}
\frac{1}{\epsilon} \,\langle I \rangle_c \;.    }
\end{array}
\end{equation}
Thus all statistical fluctuations of the transport process at $T=0$
can be extracted directly from the current. 

We conclude with the remark that these findings 
from the nonequilibrium Keldsysh approach are in agreement with results from the thermodynamic
Bethe ansatz.~\cite{saleurweiss01}

\section{From weak to strong tunneling}\label{secwtos}

\subsection{BSG or Schmid model}\label{secbsg}

There is an exact duality for the nonlinear mobility at temperature $T$ of a Brownian 
particle in a tilted periodic potential.~\cite{zwerger87,weissbook}
In the duality transformation, weak and strong corrugation of the potential are exchanged
and the spectral density is transformed.
One finds that super-Ohmic damping maps on sub-Ohmic and vice versa,~\cite{sassetti96}
while Ohmic damping maps on Ohmic (with different high-frequency cut-off), but the damping 
strength $K$ maps on $1/K$.~\cite{fisher85} The duality becomes an exact self-duality
in the Ohmic scaling limit.~\cite{schmid83} 
Independently, thermodynamic Bethe ansatz computations~\cite{fendley95b} established that 
there is an exact nonperturbative self-duality for the current in the BSG model at zero
temperature. Considering the dimensionless current ${\cal I}(K,\, \epsilon/\epsilon_0^{})
= 2\pi K I/\epsilon$ as an analytic function of $K$, gives the self-duality as
\begin{equation}\label{selfdual}
{\cal I}(K,\,\epsilon/\epsilon_0^{}) \;=\; 1 - {\cal I}(1/K,\, \epsilon/\epsilon_0^{}) \; .
\end{equation}
This equation applies for all $K>0$. Doing perturbation theory in the strong-backscattering
limit yields
\begin{equation}\label{sbslim}
{\cal I}(K,\,\epsilon/\epsilon_0^{}) = \sum_{n=1}^\infty i_n^{}(K) 
\left(\frac{\epsilon}{\epsilon_0^{}}\right)^{(2K-2)n} \; .
\end{equation}
The thermodynamic Bethe ansatz gives the coefficients $i_n^{}(K)$ in the form~\cite{fendley95b}
\begin{equation}\label{sbscoeff}
i_n^{}(K) = \frac{(-1)^{n-1}_{}}{n!}\,\frac{\Gamma(\frac{3}{2})\,\Gamma (1+nK)}{
\Gamma[\,\frac{3}{2} +n(K-1)\,]}\; .
\end{equation}
With the perturbative series (\ref{sbslim}), the self-duality (\ref{selfdual})
then yields the corresponding asymptotic or weak-backscattering expansion. 

The existence of the self-duality (\ref{selfdual}) has prompted questions about the
analytic structure of the BSG transport problem. It has been argued that, knowing 
{\em a priori} that Eq.~(\ref{selfdual}) holds,  it should be possible to find
the exact expression for ${\cal I}$ without using the Bethe ansatz. Following these lines,
exact integral representations for the current were derived which harbour both the 
perturbative and the asymptotic expansion.~\cite{fendley98,weiss96,weissbook}

Using the explicit form of the rapidity-dependent transmission coefficient in the
thermodynamic Bethe ansatz, the shot noise~\cite{fendley95a} and also the full
probability distribution $\tilde{P}(k,t)$ were derived.~\cite{saleurweiss01} The
strong-backscattering expression for the characteristic function was found to be in 
Poissonian form
\begin{eqnarray}\label{inovern}
\tilde{P}(k,t) &=& \prod_{n=1}^\infty \exp[\,t(e^{ikn}_{} - 1) \gamma_n^+ \,]\;, \\[1mm]
\gamma_n^+ &=& \frac{(-1)^{n-1}_{}}{n!} \frac{\Gamma(\frac{3}{2})\Gamma(Kn)}{ 
\Gamma[\frac{3}{2} + (K-1)n]} \frac{\epsilon}{2\pi }
\left(\frac{\epsilon}{\epsilon_0^{}}\right)^{(2K-2)n} . \nonumber
\end{eqnarray}
While a classical Poisson process requires all the $\gamma_n^{+}$ to be  positive,
the signs of the rates are alternating. The first contribution is indeed a Poisson 
process for the tunneling of electrons, but the tunneling process of a correlated pair of
electrons (and multiples thereof) has the wrong sign, which is a distinctive feature  of
quantum interference. 

The weak-backscattering expression was found as
\begin{eqnarray}\label{barinovern}
\tilde{P}(k,t) &=& {\displaystyle e^{ik\epsilon t/2\pi K}_{} \prod_{n=1}^\infty
\exp\Big[\,t(e^{-i k n/K}_{}- 1)\bar{\gamma}_n^{+}\,\Big]\; ,   }\\[1mm] \nonumber
\bar{\gamma}_n^+ &=&  {\displaystyle 
\frac{(-1)^{n-1}_{}}{n!} \frac{\Gamma(\frac{3}{2})\Gamma(\frac{n}{K})}{ 
\Gamma[\frac{3}{2} + (\frac{1}{K}-1)n]} \frac{\epsilon}{2\pi K }
\left(\frac{\epsilon}{\epsilon_0^{}}\right)^{(\frac{2}{K}-2)n} .  } 
\end{eqnarray}
The first exponential factor represents the current in the absence of tunneling.
The exponents have now the term $e^{-ikn/K}_{}$. The factor $1/K =\nu$ is the signature
of tunneling of Laughlin quasiparticles and multiples thereof. The minus sign occurs
because their tunneling diminishes the current, instead of building it up as in the 
strong-backscattering limit. The sign of the transfer weights $\bar{\gamma}_n^{+}$ is now
dependent on $K$. Using the reflection formula for gamma functions, the sign of 
$\bar{\gamma}_n^{+}$ is the one of $\cos(n\pi/K)$ and therefore positive for modest $n$
when $K\gg 1$. As $K\to \infty$, the classical regime is reached,
in which all rates $\bar{\gamma}_n^{+}$ are positive, and the fluctuations 
disappear.~\cite{saleurweiss01}
In this limit, the expression (\ref{barinovern}) is summed to the form 
\[
\tilde{P}(k,t) = \exp\left\{ ikt\frac{\epsilon}{2\pi K}
\sqrt{1 -\left(\frac{\epsilon_0^{}}{\epsilon}\right)^2}  \,\right\}\Theta(\epsilon -\epsilon_0^{}) 
\; .
\]

We see from Eq.~(\ref{inovern}) that $\gamma_n^{+}$ is exactly of order 
$x_{\rm S}^{2n}$. Hence the integrable approach is in correspondence with the findings 
(\ref{res1}) of the perturbative analysis. Conversely, the rigorous nonequilibrium 
Keldysh approach presented in Sec.~\ref{secwt} thus corroborates  
the thermodynamic Bethe ansatz. As an important by-product,
Eq.~(\ref{inovern}) provides an analytic expression for the particular
linear combination of noise integrals given in Eq.~(\ref{res1}). 

We have mentioned already at the beginning of Sec.~\ref{secwt} that the noise integrals
${\cal J}_\ell^{}(\vec{\xi}\,)$ diverge at short distances for $K\ge \frac{1}{2}$. 
At this point two remarks are appropriate. 
(i) It is most natural to define the regularized integrals for $K \ge \frac{1}{2}$ as the
analytic continuation of their values for $K < \frac{1}{2}$. Since the integrals
have simple poles and there are no branch points, the analytic continuation is 
well-defined. (ii) The friction factors exhibit zeros at the locations of the poles of 
the integrals.
As a result, the expression (\ref{res1}), in which friction factors and noise integrals 
are combined, is a smooth function of $K$ for all $K$,
as can be seen from expression (\ref{inovern}). Also the Kondo scale for the BSG model,
Eq.~(\ref{kondoscale}), is nonsingular for all $K$. On the other hand, the remaining 
singularities in the SB expression (\ref{uwrel}) 
can be fully absorbed into the definition of the Kondo scale for the SB model
given in Eq.~(\ref{kondoscale}), as we see from relation (\ref{raterel}).
From this we may conclude that the rate expressions found from our analysis for the Schmid and 
SB model are valid for all $K>0$. 

\subsection{Spin-boson model}\label{secspinboson}

With use of relation (\ref{raterel}) and expression (\ref{inovern}) for the
BSG rate $\gamma_n^{+}$, we immediately obtain an exact expression in closed
analytic form for the partial tunneling rate $\tilde{\gamma}_n^{+}$ in  the spin-boson model.
With this, the weak-tunneling expansion of the full rate $\tilde{\gamma}_{}^{+}$, 
Eq.~(\ref{wtexpan}), is found to read
\begin{equation}\label{series3}
\begin{array}{rcl}
\tilde{\gamma}^+ &=& {\displaystyle \frac{\epsilon}{2\sqrt{\pi}}\,
 \sum_{m=1}^{\infty }  a_m^{}(K)\,
\left(\frac{\epsilon}{\epsilon_0^{}}\right)^{(2K-2)m}_{} \; , }\\[6mm] 
a_m^{}(K) &=&
{\displaystyle \frac{1}{m!}\, \frac{ \Gamma (Km)\,[\,1- \cos (2\pi K m)\,]}{
\Gamma[\,\frac{3}{2} +(K-1)m\,]}} \,   \; . 
\end{array}
\end{equation}
For rational $K$, the series (\ref{series3}) can be written
as a linear combination of hypergeometric functions.

In the regime $K<1$, the perturbative series (\ref{series3}) is absolutely converging
for large enough bias. For $K>1$, the series 
is absolutely converging for small enough bias.
The leading term is the golden rule rate \cite{leggett87,weissbook}
\begin{equation}\label{goldrule}
\begin{array}{rcl}
\tilde{\gamma}_{\rm GR}^+ 
 &=& {\displaystyle \frac{\pi}{2\,\Gamma(2K)}\,\epsilon\, x_{\rm SB}^{2} } \\[4mm]
 &=& {\displaystyle 
\frac{\sin^2(\pi K)\Gamma(K)}{\sqrt{\pi}\Gamma(\frac{1}{2} + K)}\,\,\epsilon
\left(\frac{\epsilon}{\epsilon_0^{}}\right)^{2K-2} \; . }
\end{array}
\end{equation} 
The coefficients $a_m^{}(K=\frac{1}{2})$ are zero when $m\neq 1$.
Hence $\gamma_{\rm GR}^{+}$ coincides with the full rate for $K= \frac{1}{2}$.

An integral representation for $\tilde{\gamma}^+$ is found by 
writing the quotient of the Gamma functions in Eq.~(\ref{series3}) as a contour
integral,\cite{temme}
\begin{equation}\label{contint}
\frac{\Gamma(x)}{\Gamma(x+y)}
=  \frac{\Gamma(1-y)}{2\pi i}\int_{\cal C} {\rm d}z\,
 z^{x-1}_{}(z-1)^{y-1}_{} \; .
\end{equation}
The contour starts at the origin, circles around the branch point in 
counter-clockwise sense, and returns to the origin.
Equating $x$ with $Km$ and $y$ with $\frac{3}{2} - m $, and putting
$u_1^{}=(\epsilon/\epsilon_0^{})^{2K-2}_{}$ and 
$u_2^{}= e^{i 2\pi K}_{} u_1^{}$, we get
\begin{equation}
\begin{array}{rcl}
\tilde{\gamma}^{+}_{} &=& {\displaystyle \frac{\epsilon}{2\sqrt{\pi}}
\sum_{m=1}^\infty \frac{\Gamma(m -\frac{1}{2})}{m!}\;{\rm Re}\Big\{\, u_1^m - u_2^m\,
\Big\}}   \\[5mm]
&&\times\;{\displaystyle \frac{1}{2\pi i}\int_{\cal C}{\rm d}z\,
\frac{\sqrt{z-1}}{z}\left(\frac{z^{K}_{}}{z-1}\right)^m } \; .
\end{array}
\end{equation}
Since the weak-tunneling series (\ref{series3}) is absolutely convergent 
within the circle of 
convergence, we can interchange the order of integration and summation.
This yields
\begin{equation}\label{intrep1}
\begin{array}{rcl}
\tilde{\gamma}^+ &=& {\displaystyle {\rm Re}\,\frac{\epsilon}{2\pi i}\!\!
\int_{\cal C}^{} \! \frac{{\rm d}z}{z} } \\[4mm]
&&\times {\displaystyle
\left[\sqrt{z-1 -z^K_{}u_2^{}}
\,-\, \sqrt{z-1 -z^K_{}u_1^{}}\,\right]}  \; .
\end{array}
\end{equation}

The integral representation (\ref{intrep1}) converges for all $K$ and all 
$\epsilon/\epsilon_0^{}$, not just in the 
regime where the series (\ref{series3}) does. Evidently, Eq.~(\ref{intrep1}) is 
generally valid. Inter alia, it must also hold 
the corresponding strong-tunneling or asymptotic expansion for small (large) enough bias 
when $K<1$ ($K > 1$).

Consider first the asymptotic expansion for $K < 1$. Upon changing variable $z$
to $y= z^{1-K}_{}/u_1^{}$ and $y= z^{1-K}_{}/u_2^{}$, respectively, 
Eq.~(\ref{intrep1}) transforms into 
\begin{equation}\label{intrep2}
\begin{array}{rcl}
\tilde{\gamma}_{}^+ &=& {\displaystyle
\frac{1}{1-K}\; \frac{\epsilon_0^{}}{2\pi i} \int_{\cal C}
\frac{{\rm d}y}{y} \; y^{\frac{K}{2(1-K)}_{} }}\\[6mm]
&\times&\!\! {\rm Re}\,{\displaystyle \left\{   e^{i\pi\frac{K}{1-K}}_{}
\sqrt{y - 1 -\Big(y e^{i2\pi}_{}\Big)^{\frac{K}{K-1}}
\left(\frac{\epsilon}{\epsilon_0^{}}\right)^2 }   \right. }\\[6mm]
&&\qquad\quad\quad\;\; -\; {\displaystyle \left. 
\sqrt{y - 1 - y^{\frac{K}{K-1}}_{}\left(\frac{\epsilon}{\epsilon_0^{}} 
\right)^2} \;\right\} }   \,.
\end{array}
\end{equation}
Next we expand the curly bracket for small $\epsilon/\epsilon_0^{}$ 
and interchange summation and integration. Using Eq.~(\ref{contint}), the 
strong-tunneling series is found as
\begin{equation}\label{asymp1}
\tilde{\gamma}^+ = \sum_{n=0}^\infty  \tilde{\gamma}_n^+ \; ,\qquad
 \tilde{\gamma}_n^+ =
\frac{\epsilon_0^{}}{2\sqrt{\pi} } \, b_n^{}(K) 
\,\left(\frac{\epsilon}{\epsilon_0^{}}\right)^{2n}_{} \; .
\end{equation}
For $K\le \frac{1}{3}$, the coefficient $b_n^{}(K)$ is given by
\begin{equation}\label{as2}
b_n^{}(K) = {\textstyle 2\sin^2[\frac{\pi K}{1-K}(\frac{1}{2}-n)]
\,c_n^{}(K) \;, }
\end{equation}
where
\begin{equation}\label{asymp2}
c_n^{}(K) =
\frac{1}{ n!}\,\frac{\Gamma[(\frac{1}{2}-n)
\frac{K}{1-K}]}{(\frac{1}{2}-n)\Gamma[(\frac{1}{2}-n)\frac{1}{1-K}]} \;.
\end{equation}
For $ \frac{1}{3} < K < 1 $, the contour does not circle the branch
cut of the first term in the curly bracket in Eq.~(\ref{intrep2}). Hence only the second
term contributes and yields
\begin{equation}\label{as1} 
b_n^{}(K) = c_n^{}(K) \; .
\end{equation}

The strong-tunneling or asymptotic series (\ref{asymp1}) is an expansion in powers of
$\epsilon^2_{}/\epsilon_0^2$. The leading term represents the forward rate of the 
symmetric SB model,
\begin{equation}\label{gamma0}
\tilde{\gamma}_0^+ = \frac{A_K^{}}{\sqrt{\pi}}\,
\frac{\Gamma[\frac{K}{2(1-K)}]}{\Gamma[\frac{1}{2(1-K)}]}\,\epsilon_0^{}\; , 
\end{equation}
where $A_K^{}= 2 \sin^2[\pi K/2(1-K)]$ for $K \le \frac{1}{3}$, and
$A_K^{}=1$ for $\frac{1}{3} < K < 1$.

Consider next the weak-damping limit. For  $K  \ll 1$, the perturbative
series (\ref{series3}) as well as the asymptotic series (\ref{asymp1})
can be summed to the simple form
\begin{equation}
\tilde{\gamma}^+_{} = \pi K \Delta_{\rm SB}^{2}\Big/
\sqrt{\Delta_{\rm SB}^2 +\epsilon^2_{}} \; ,
\end{equation} 
which is in agreement with findings from a systematic weak-damping 
self-energy method.~\cite{woll89,weissbook}

\begin{figure}[t]
\includegraphics[width=80mm]{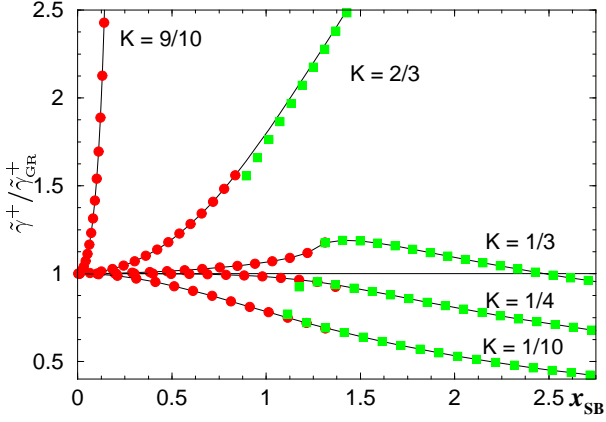}
\caption{The scaled rate
$\tilde{\gamma}^+_{}/\tilde{\gamma}_{\rm GR}^+$ is plotted versus
$x_{\rm SB}^{}$ for various $K<1$.
The circles represent the weak-tunneling
series and the squares the strong-tunneling expansion. The full curve is
the hypergeometric function expression
\label{figone}}
\end{figure}

In Fig.~\ref{figone}, the normalized rate 
$ \tilde{\gamma}^+_{}/ \tilde{\gamma}^+_{\rm GR}$ is shown for
different $K < 1$. The horizontal line represents the particular case $K=\frac{1}{2}$.
For $K < \frac{1}{4}$, the full rate is always lower than the golden rule rate. Hence the
numerous tunneling paths interfere destructively in this regime.
For $\frac{1}{2} < K < 1$, the tunneling contributions interfere
constructively 
for all $x_{\rm SB}^{}$ so that the full rate is always above the
golden rule rate. In the regime $\frac{1}{4} < K < \frac{1}{2}$, the normalized
rate goes through a maximum as tunneling is increased, and finally falls below one.  
This reflects constructive interference at small
and intermediate $x_{\rm SB}^{}$, and destructive interference at large
$x_{\rm SB}^{}$.

\begin{figure}[t]
\includegraphics[width=80mm]{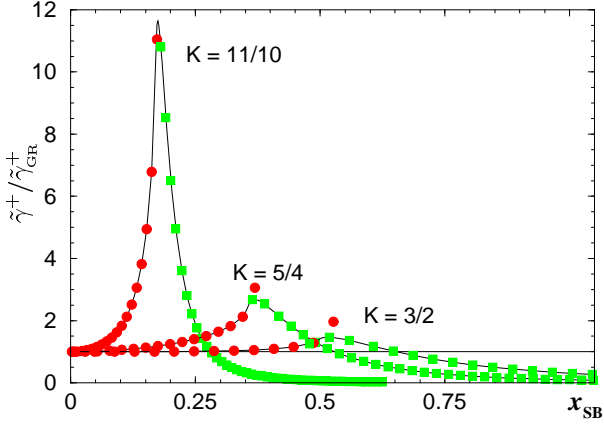}
\caption{The same scaled rate, but now for various $K > 1$. \label{figtwo}}
\end{figure}

Next consider large damping $K>1$ and large bias.
It is useful to divide the damping regime into the sections
$K= p + \kappa$ with $p=1,2,\cdots$ and $0 \le \kappa < 1$.
In the first step, we change in the integral representation
(\ref{intrep1}) from
variable $z$ to variable $t = e^{-i\pi}_{}u_{1/2}^{}z^K_{}$. This gives
\begin{equation}
\tilde{\gamma}^{+} = {\rm Re}\,\frac{1}{K} \,\frac{\epsilon}{2\pi i}
\int_{\cal C} {\rm d}t\,\frac{H(t,u_1^{})}{t} \;,
\end{equation}  
\begin{equation}
\begin{array}{rcl}
H(t,u_1^{})&=&  {\displaystyle \sqrt{ t - 1
+e^{i\pi(1+2p)/K}_{}(u_1^{}t)^{1/K}_{}} }    \\[4mm]  
&&\quad\quad \quad  -\; {\displaystyle \sqrt{t-1 + e^{i \pi/K}_{}(u_1^{}t)^{1/K}_{}}\;  \; . } 
\end{array}
\end{equation}
Second, we expand $H(t,u_1^{}$ for small $u_1^{}$ and interchange
integration and summation. Using (\ref{contint}), we then obtain the
asymptotic series as
\begin{eqnarray}\label{asymp3}
\tilde{\gamma}^+ &=& \frac{\epsilon}{2\sqrt{\pi} }\, 
\sum_{m=1}^\infty d_m^{}(K)\, 
\left(\frac{\epsilon}{\epsilon_0^{}}\right)^{(2/K - 2)m}_{}  \; , \\[2mm]  \nonumber
d_m^{}(K) &=& 
\frac{(-1)^{m}_{}}{m!}\,\frac{2\, \Gamma(\frac{m}{K})
\sin[\,\frac{1+p}{K} m\pi] 
\sin(\frac{p}{K}m\pi)}{ K\, \Gamma[\,\frac{3}{2} 
+  (\frac{1}{K} -1 )m\, ]} \; .
\end{eqnarray}
The asymptotic series (\ref{asymp3}) for the SB model in the regime $K>1$
bears a strong resemblance with the corresponding series of 
the self-dual Schmid model (\ref{barinovern}).
The powers of $\epsilon/\epsilon_0^{}$ follow from those 
of the perturbative expansion (\ref{series3}) by the substitution $K\to 1/K$, 
as also do major parts of the coefficient $d_m^{}(K)$. 
However, complete duality is spoilt by the alternating sign and by the
sine-factors.

Fig.~\ref{figtwo} shows plots of the normalized rate 
$ \tilde{\gamma}^+_{}/ \tilde{\gamma}^+_{\rm GR}$
for $K>1$. The normalized rate goes through a maximum which is shifted to higher 
$x_{\rm SB}^{}$ when $K$ is increased. At large enough $x_{\rm SB}^{}$,
the rate $\tilde{\gamma}^+_{}$ falls below the golden rule rate. Hence there is constructive
interference of tunneling at small and intermediate $x_{\rm SB}^{}$,
and destructive interference in the large tunneling regime.

\subsection{Conjecture on decoherence at $T=0$}\label{secconj}

So far, we have left aside the damping of the coherent oscillations of the populations.
If one aims at calculating frequency and decoherence
rate of the oscillation directly, the $\lambda$-dependence of the
irreducible clusters discussed above has to be taken into account. 
The calculation of the respective self-energy and of the decoherence rate has been 
carried out in special limits,~\cite{weissbook} e.g. for $K\ll 1$.~\cite{woll89}
Here we pose the conjecture that the strong tunneling expansion of the
tunneling relaxation rate $\tilde{\gamma}^{+}_{}$, Eq.~(\ref{asymp1}),
and the corresponding expansion of the decoherence rate
\begin{equation}
\gamma^{\rm (d)}_{} = \sum _{n=0}^\infty \gamma_n^{\rm (d)}\; ;\qquad
\gamma_n^{\rm (d)} = \frac{\epsilon_0^{}}{2\sqrt{\pi}}\, f_n^{}(K) 
\left( \frac{\epsilon}{\epsilon_0^{}}\right)^{2n}
\end{equation}
are closely related, viz. that the ratio of the
coefficients of these expansions is given by
\begin{equation} \label{rela}
\frac{f_n^{}(K)}{ b_n^{}(K)} = \left\{ 
\begin{array}{ll}
\frac{1}{2}  \; , 
      &  K \le \frac{1}{3} \;, \\[3mm] 
\sin^2[\frac{\pi K(1-2n)}{2(1-K)} ] \;,
\quad    &   \frac{1}{3} < K < \frac{1}{2} \; .\\
\end{array}
\right.
\end{equation}
Equation~(\ref{rela}) is in agreement with all exact expressions known in special limits.

The decoherence rate of the unbiased  SB model has been calculated 
within the framework of integrable QFT.~\cite{lesage98}
The respective result (translated into our notation) is
\begin{equation}
\gamma_0^{\rm (d)} = \frac{1}{\sqrt{\pi}} {\textstyle
\sin^2_{}[\frac{\pi K}{2(1-K)}]}\,\frac{\textstyle 
\Gamma[\frac{K}{2(1-K)}]}{\textstyle
\Gamma[\frac{1}{2(1-K)}]}\, \epsilon_0^{}  
\end{equation}
is in correspondence with expressions (\ref{rela}) and (\ref{gamma0}) for 
$K\le \frac{1}{2}$. The relation (\ref{rela}) is known to hold also in the
biased case in the regimes $K\ll 1$ and $K$ close to
$\frac{1}{2}$.~\cite{weissbook} 

These checks suggest that the conjecture (\ref{rela}) holds for the SB model at zero temperature
in the regime $0< K  \le \frac{1}{2}$. Upon combining
Eqs.~(\ref{asymp1}) - (\ref{asymp2}) with (\ref{rela}) 
the weak-bias expansion of the decoherence rate is found as 
\begin{equation}\label{deco}
\gamma_{}^{\rm (d)} = \frac{\epsilon_0^{}}{2\sqrt{\pi}}\sum_{n=0}^\infty
{\textstyle \sin^2[\frac{\pi K}{1-K}(\frac{1}{2}-n)]}
\,c_n^{}(K) \left(\frac{\epsilon}{\epsilon_0^{}}\right)^{2n} \!\! .
\end{equation}
If the conjecture (\ref{rela}) is correct, then the series expression (\ref{deco}) 
represents the lower bound for decoherence in the SB model in the scaling limit
in the regime  $0 <K \le \frac{1}{2}$.

\subsection{Leading enhancement at low temperatures}\label{subsecenhanc}

We now extend the discussion to the asymptotic low-temperature regime. The leading thermal 
correction in the noise amplitude factor $G_\ell^{}$ is found from Eq.~(\ref{noiseampl}) as
\[
G_{\ell}^{} = G_{\ell}^{}(\vartheta=0)
\Big\{ 1 - K \frac{\pi^2_{}}{3}\,\vartheta^2_{} 
\Big(\sum_{j=1}^{2\ell-1}p_j^{}\tau_j^{}\Big)^2 +{\cal O}(\vartheta^4_{})
\Big\} \;,
\]
The important point now is to recognize that
the annoying term in the round bracket is the bias phase (\ref{phaserel}). Therefore, 
this term can be generated by differentiation of the noise integral (\ref{noiseint})
with respect to the bias.~\cite{sass90} We thus have, for example,
\begin{equation}\label{enhanc}
\gamma_n^{+}(\epsilon,T) = \left\{ 1  + K \frac{\pi^2_{}}{3}\,T^2_{} 
\frac{\partial^2_{}}{(\partial \epsilon)^2_{}}\right\} \gamma_n^{+}(\epsilon,\,T=0) \; .
\end{equation}
Corresponding expressions hold for the current $\langle I \rangle_c^{}$ and for the moments
$\langle I^{(m)}_{}\rangle_c^{}$ given in Eq.~(\ref{curmom}) both in the weak- and strong
tunneling regime.
Equally, the thermal enhancement of the tunneling rate and the decoherence rate in the SB model 
has a form analogous to Eq.~(\ref{enhanc}).

We see that the leading enhancement at $T\ll \epsilon$ varies with $T^2_{}$.
The power 2 is a signature of Ohmic dissipation.
In addition the prefactor is universally given by the curly bracket with the
second order differential expression in Eq.~(\ref{enhanc}). The universal form is closely 
related to the Wilson ratio in the AKM, in which the specific heat is related to the static
susceptibility.~\cite{sass90}
The physical origin of the $T^{2}_{}$ enhancement is the low-frequency thermal noise of 
the dissipation.\cite{grabert88}
\vspace{5mm}

\section{Conclusion}\label{concl}

In summary, we uncovered exact functional relations between perturbative integrals
occuring in the nonequilibrium Keldysh approach applied to the Schmid and the SB model at 
$T=0$. With that and with use of the correspondence of the Schmid with the BSG model,
we found agreement between the results of the Keldysh approach and the thermodynamic 
Bethe ansatz for the BSG model.
As a second result, we discovered exact functional
relations between the rates in the Schmid model and the
partial rates in the SB model. Starting out from the perturbative series obtained by use of the 
functional relations, we put up an exact integral representation
for the tunneling rate in the SB model, from where we derived the asymptotic or
strong-backscattering expansion. Thirdly, we conjectured  and checked a relation
between the relaxation and the decoherence rate in the SB model.
Finally, we calculated the leading thermal enhancement of the transport and of
all statistical fluctuations and found universal behaviour.
\vspace{1mm}

Our results  are relevant, e.g., to charge transfer, 
macroscopic quantum coherence (qubits), and diverse quantum impurity
problems.


\begin{thebibliography}{199}


\bibitem{weissbook}
U. Weiss, {\em Quantum Dissipative Systems} (World Scientific,
Singapore, 2nd edition, 1999).

\bibitem{fendley95a}
P. Fendley and H. Saleur, Phys. Rev. Lett. {\bf 75}, 4492 (1995). 

\bibitem{fendley95c}
P. Fendley, F. Lesage, and H. Saleur, J. Stat. Phys. {\bf 85}, 211 (1996). 

\bibitem{leggett87}
A.J.~Leggett {\it et al.},
Rev. Mod. Phys. {\bf 59}, 1 (1987).

\bibitem{schmid83}
A.~Schmid, Phys. Rev. Lett. {\bf 51}, 1506 (1983).

\bibitem{picciotto}
R. de-Picciotto {\it et al.}, Nature {\bf 389}, 162 (1997).

\bibitem{roddaro}
S.~Roddaro {\it et al.}, Phys. Rev. Lett. {\bf 90}, 46805 (2003).

\bibitem{kondo88}
J.~Kondo, in {\em Fermi Surface Effects}, 
Vol.~77 of Springer Series in Solid State Sciences, eds. J.~Kondo and A.~Yoshimori
(Springer, Berlin, 1988).

\bibitem{tsvelik}
A.M. Tsvelik and P.B. Wiegmann, Adv. Physics, {\bf 32}, 453 (1983).

\bibitem{fendley95b}
P. Fendley, A. Ludwig, and H. Saleur, Phys. Rev. B {\bf 52}, 8934 (1995).

\bibitem{saleurweiss01} 
H. Saleur and U. Weiss, Phys. Rev. B {\bf 63} 201302(R) (2001).

\bibitem{zwerger87} W. Zwerger, Phys. Rev. B {\bf 35}, 4737 (1987).

\bibitem{sassetti96} M. Sassetti, H. Schomerus, and U. Weiss, Phys. Rev. B~{\bf 53},
R2914 (1996).

\bibitem{fisher85}
M.P.A. Fisher and W. Zwerger, Phys. Rev. B {\bf 32}, 6190 (1985).

\bibitem{weiss96}
U.~Weiss, Solid. State Comm. {\bf 100}, 281 (1996).

\bibitem{fendley98}
P. Fendley and H. Saleur, Phys. Rev. Lett. {\bf 81}, 2518 (1998).

\bibitem{temme}
N.M. Temme, {\it Special Functions}, (Wiley, New York, 1996).

\bibitem{woll89}
U. Weiss and M. Wollensak, Phys. Rev. Lett. {\bf 62}, 1663 (1989).

\bibitem{lesage98}
F. Lesage and H. Saleur, Phys. Rev. Lett. {\bf 80}, 4370 (1998). 

\bibitem{sass90} M. Sassetti and U. Weiss, Phys. Rev. Lett. {\bf 65}, 2262 (1990).

\bibitem{grabert88} J.M. Martinis and H. Grabert, Phys. Rev. B {\bf 38}, 2371 (1988).

\end{thebibliography}
\end{document}